\documentclass[letterpaper,journal]{IEEEtran}
\usepackage{amsmath,amssymb,amsfonts,mathrsfs,mathtools,bm, bbm, dsfont,mathrsfs,amsthm,blkarray}
\usepackage[dvipdfm]{graphicx}
\usepackage[dvipsnames]{xcolor}
\usepackage{graphicx}
\usepackage[margin=0.7in]{geometry}
\usepackage{multirow}
\usepackage{epsfig,epsf,psfrag,latexsym}
\usepackage{xurl}
\usepackage{booktabs}
\usepackage{subcaption} 
\usepackage{comment} 
\usepackage{graphicx}
\usepackage[ruled,vlined]{algorithm2e}
\usepackage{amsmath,amssymb,amsthm}

\usepackage{booktabs}
\usepackage{multirow,multicol}

\def\D{{\mbox{\rm\tiny D}}}
\def\U{{\mbox{\rm\tiny U}}}
\def\R{{\mbox{\rm\tiny R}}}

\def\TLMP{{\mbox{\rm\tiny TLMP}}}
\def\mTLMP{{\mbox{\rm\tiny MTLMP}}}
\def\mDP{{\mbox{\rm\tiny MDCP}}}

\usepackage{booktabs}
\usepackage{array}
\usepackage{xcolor}
\usepackage[breaklinks,colorlinks, linkcolor=MidnightBlue, anchorcolor=MidnightBlue, citecolor=MidnightBlue, urlcolor=MidnightBlue]{hyperref}
\usepackage{cite}

 \def\old#1{}    % Please don't remove this... This command includes the text to be deleted.

\newcommand{\tcb}[1]{\textcolor{blue}{#1}} % VN: I will write with blue to make visible what I wrote.

\colorlet{blue}{black}

\def\nn{\nonumber}
\def\beq{\begin{equation}}
\def\eeq{\end{equation}}
\def\bea{\begin{eqnarray}}
\def\eea{\end{eqnarray}}
\def\ba{\begin{array}}
\def\ea{\end{array}}

\def\bitem{\begin{itemize}}
\def\eitem{\end{itemize}}
\def\ben{\begin{enumerate}}
\def\een{\end{enumerate}}

\def\ie{{\it i.e.,\ \/}}

\def\tcb{\textcolor{blue}}

\def\mubf{\hbox{\boldmath$\mu$\unboldmath}}

\def\pibf{\hbox{\boldmath$\pi$\unboldmath}}

\def\Sigmabf{\hbox{$\bf \Sigma$}}

\def\cbf{{\bm c}}
\def\dbf{{\bm d}}

\def\gbf{{\bm g}}

\def\lbf{{\bm l}}

\def\rbf{{\bm r}}
\def\sbf{{\bm s}}

\def\xbf{{\bm x}}
\def\ybf{{\bm y}}

\def\rbf{{\bm r}}
\def\xbf{{\bm x}}
\def\ybf{{\bm y}}

\def\Nc{{\cal N}}

\newcommand{\beqa}{\begin{eqnarray}}
\newcommand{\eeqa}{\end{eqnarray}}
\newcommand{\beqan}{\begin{eqnarray*}}
\newcommand{\eeqan}{\end{eqnarray*}}

\renewcommand{\arraystretch}{0.9}

\renewcommand{\arraystretch}{1.1}

\renewcommand{\[}{\left[}

\newcounter{l1}
\newcounter{l2}
\newcounter{l3}
\newcommand{\bdotlist}{\begin{list}{$\bullet$}{}}
\newcommand{\bboxlist}{\begin{list}{$\Box$}{}}
\newcommand{\bbboxlist}{\begin{list}{\raisebox{.005in}{{\tiny
$\blacksquare$ \ \ }}}{}}
\newcommand{\bdashlist}{\begin{list}{$-$}{} }
\newcommand{\blist}{\begin{list}{}{} }
\newcommand{\barablist}{\begin{list}{\arabic{l1}}{\usecounter{l1}}}
\newcommand{\balphlist}{\begin{list}{(\alph{l2})}{\usecounter{l2}}}
\newcommand{\bAlphlist}{\begin{list}{\Alph{l2}.}{\usecounter{l2}}}
\newcommand{\bdiamlist}{\begin{list}{$\diamond$}{}}
\newcommand{\bromalist}{\begin{list}{(\roman{l3})}{\usecounter{l3}}}

\newtheorem{theorem}{Theorem}

\newtheorem{proposition}{Proposition}

\title{Ramping Procurement and Bid-Cost Recovery in Real-Time Market}
\author{

Cong Chen, ~\IEEEmembership{Member,~IEEE,}
\quad Valentina  Norambuena,~\IEEEmembership{Student Member,~IEEE,}
\quad Lang~Tong, ~\IEEEmembership{Fellow,~IEEE}
\thanks{\scriptsize Part of the work was accepted by the 2025 IEEE Power \& Energy Society General Meeting  (PESGM) \cite{ChenTong25PESGM}.}
\thanks{\scriptsize
Cong Chen (Cong.Chen@dartmouth.edu) is with Thayer School of Engineering, Dartmouth College, Hanover, NH, USA. Valentina  Norambuena and Lang Tong are with the School of Electrical and Computer Engineering, Cornell University, Ithaca, NY, USA. This work is supported in part by the National Science Foundation under Award 2218110 and 2412776, and Power Systems and Engineering Research Center (PSERC) Research Project M-46.}
}

\begin{document}
% \vspace{-cm}
\include{pythonlisting}

\maketitle
\begin{abstract}

We study ramping procurement co-optimized with economic dispatch under net-demand uncertainty. We examine two flexible ramp product designs implemented by grid operators: single-interval and multi-interval co-optimization. Both rely on rolling-window stochastic optimization with binding and advisory interval decisions. We develop analytical frameworks to evaluate generator profits, consumer payments, bid cost recovery (BCR), and operational efficiency.  In particular, net-demand uncertainty may lead to generator under-compensation, requiring discriminatory BCR. While operational efficiency is invariant to energy and ramp prices, producer profits and consumer payments depend critically on pricing. We examine locational marginal pricing (LMP)  and two uniform pricing: maximum dispatch cost pricing (MDCP) and maximum temporal locational marginal pricing (MTLMP). With out-of-market BCR, LMP yields discriminatory energy prices, whereas MDCP eliminates BCR and MTLMP does so in most cases. This property enables us to establish truthful bidding incentives for price-taking generators under MDCP. \textcolor{blue}{Our analysis highlights trade-offs between single- and multi-interval co-optimization and pricing designs: multi-interval energy-ramp co-optimization outperforms single-interval co-optimization when ramping forecasts are accurate and ramping needs are tight, but its advantage diminishes under large forecast errors and moderate ramping needs.} Empirical results on CAISO and ERCOT data show that MDCP and MTLMP increase producer profits with negligible BCR, albeit at the expense of higher consumer payments relative to LMP.
\end{abstract}


%\vspace{-0.4cm}
\section{Introduction}\label{sec:Intro}
Deep penetration of renewable energy resources has increased net-load uncertainty and the need for both upward and downward ramping capability in the real-time market. This challenge is particularly pronounced in  renewable-rich regions such as those operated by the California  Independent System Operator (CAISO) and Midcontinent Independent System Operator (MISO). The substantially increased ramping needs and inadequate ramping support has resulted in the excess use of system reserves and increasing renewable curtailments \cite{Clyde24CAISOramp, MISO:Schedule29}. Since 2016, both MISO and CAISO have developed  Flexible Ramp Product or Ramp Capability Product (herein referred to as FRP), aimed at positioning generators to provide ramping supports via energy and ramp-capacity co-optimizations \cite{cavicchi18ramp}.

Significant challenges remain, however. First, because existing FRP implementations do not use bid-based ramp procurement and the ramp demand curves are set administratively, ramp prices are often too low under current Locational Marginal Pricing (LMP)-based energy and ramp pricing \cite{cavicchi18ramp,FERC15upliftE-2_14}. Thus, LMP and ramp capacity shadow price may not be sufficiently high to incentivize generation and storage resources to provide ramping support in real-time dispatch.  

Second, to procure ramping capacity to meet anticipated ramping needs in upcoming hours, generators may need to forgo profitable real-time energy opportunities. As a result, energy LMPs may fall below generators' bid-in costs \cite{CAISO:25BCRMIO}, leading to under-compensation that must be addressed through out-of-market uplift payments. 

One of the common out-of-market uplift settlements is the make-whole payment (MWP), also known as real-time bid cost recovery (BCR), which compensates underpaid generators for their losses.\footnote{BCR provides uplifts when market revenues fail to cover a resource’s start-up, minimum-load, or energy-bid costs over the course of a day. We focus on real-time BCR/MWP  and energy-bid costs.} However, even though LMP is non-discriminative,  the uplift-adjusted energy price is discriminative and the underlying payment mechanism non-transparent. More significantly, perhaps, is that the out-of-market settlement incentivizes strategic bidding behavior that withholds capacity to gain higher profit from out-of-market BCR \cite{CAISO:22BCR}.

This paper establishes a framework for analyzing ramping procurement in the real-time electricity market, identifies the causes of out-of-market BCR uplifts, \tcb{and proposes uniform pricing mechanisms that compensate generators providing ramping capability without out-of-market uplift.} Prior work by Cavicchi and Harvey \cite{cavicchi18ramp} provides a comprehensive overview of industry practices for ramp capability dispatch and pricing. Building on their work and multiple industry reports in \cite{Clyde24CAISOramp, MISO:Schedule29, CAISO:23FRP}, we develop concrete  models for energy-ramping co-optimization and present both analytical and empirical studies that examine price volatility, out-of-market uplift payments, generator profits, demand payments, and   efficiency.

\vspace*{-10 pt}%
%\vspace{-0.4cm}
\subsection{Related work}

Ramping capability procurement serves two purposes \cite{cavicchi18ramp}: (i) accommodating intertemporal net-load changes and (ii) reserving ramping headroom to hedge against forecast errors. The former motivates {\em multi-interval rolling-window dispatch}, while the latter motivates   FRPs. Together, FRP and  rolling-window dispatch position resources in advance for anticipated ramping events under real-time operational uncertainty. %We summarize related literature in these two categories.%, for example by withholding generation capacity or charging storage ahead of future ramp-up conditions.

For FRP design, prior research has focused on the setting of ramping requirements and on the relationship between FRP and multi-interval dispatch. It has been recognized that FRP are generally not equivalent to multi-interval rolling-window dispatch. In \cite{Zhang26Ramp}, the authors establish equivalence between FRP and multi-interval dispatch in a one-shot optimization with perfect foresight, and prove that such equivalence does not hold under rolling-window dispatch with imperfect forecasts and limited look-ahead horizons. Other studies focus one day ahead operation with FRP \cite{GhaljeheiKhorsand22ADAFRP, YurdakulEla25FRPDA} and analyze the impact of ramping requirement settings on system performance \cite{wuhug15TPSfrp}. In practice, recent enhancements to FRP address network congestion and the choice of look-ahead horizons  \cite{CAISO:23FRP, cavicchi18ramp}.

For {\em  multi-interval rolling-window  dispatch} in real-time market, the literature progresses along two main directions. The first focuses on operational efficiency under uncertainty using model predictive control (MPC)-style formulations. Studies compare single-interval and multi-interval dispatch and examine the impact of forecast uncertainty on real-time operations \cite{Biggar:22EJ, ela15TPSrtFRP}. To further enhance performance under uncertainty, stochastic optimization methods have been proposed \cite{Cho:23OR, Werner23CDCpricing}. These works demonstrate efficiency gains for generators with intertemporal ramping constraints \cite{ela15TPSrtFRP} and  storage  with intertemporal state-of-charge constraints \cite{Zhao&Zheng&Litvinov:19TPS}.

\textcolor{blue}{The second line of research focuses on out-of-market uplift payments and out-of-merit dispatch. Out-of-merit dispatch refers to resources being dispatched at prices below their bid-in costs \cite{CAISO:25BCRMIO}, which results in under-compensation \cite{Hogan:20, Mays:24EE}. Such out-of-merit dispatch arises from multiple sources, including unit commitment for fast-start resources \cite{FERC15upliftE-2_14, wanghobbs15TPSfrp} and ramping limits and forecast uncertainty in rolling-window dispatch \cite{ZhangKory19rampuplift, Cho:23OR}. To compensate generators, system operators rely on out-of-market uplifts—such as lost opportunity cost (LOC) and MWP/BCR.\footnote{LOC compensates generators for their individual ex-post optimal profit—a stronger form of uplifts than MWP/BCR, which only cover energy-bid costs.} While these mechanisms correct ex-post profit shortfalls, they raise concerns regarding transparency and strategic behavior \cite{CAISO:25BCRMIO}. Consequently, substantial efforts have been devoted to developing alternative real-time pricing methods that reduce uplifts \cite{Hogan:20, Zhao&Zheng&Litvinov:19TPS, Chen&Guo&Tong:20TPS}. Prior work shows that LOC uplifts are unavoidable under uniform in-market pricing \cite{Guo&Chen&Tong:21TPS}, whereas MWP uplifts can be eliminated under certain uniform pricing mechanisms \cite{ChenTong25PESGM}.}

Despite extensive research on  FRP and {\em  multi-interval rolling-window  dispatch}, few studies jointly examine operational efficiency, out-of-merit dispatch, and out-of-market compensation in  energy-ramping co-optimization. FRPs are typically compensated through opportunity-cost-based pricing, under the assumption that generators awarded ramping opportunity costs can recover foregone energy revenues \cite{YurdakulEla25FRPDA}. While this assumption holds under perfect foresight, it breaks down in rolling-window real-time market with forecast errors and limited look-ahead horizons. This paper fills this gap by demonstrating that existing  schemes for energy and FRPs can still lead to out-of-merit dispatch and out-of-market uplifts, even when FRP payments are applied. We further propose pricing alternatives and analyze operational efficiency across  FRP implementations in single- and multi-interval dispatch.

\subsection{Summary of contribution}

This paper provides both analytical and empirical evaluations of real-time market with ramping procurement through energy–ramping co-optimization. We examine the mechanisms leading to out-of-market uplifts and investigate uniform pricing for procuring ramping capacity without BCR/MWP.

\textit{Analytical contributions:} First, we develop an energy-ramping co-optimization framework that captures the essential features of practical FRPs, from which we characterize fundamental properties of FRP dispatch and pricing. \tcb{Second, while it is intuitive that sufficiently high energy prices can eliminate MWPs, the least-cost uniform pricing mechanism that eliminates MWPs for all generators remains unknown.} We derive two uniform pricing—maximum dispatch cost pricing (MDCP) and maximum temporal locational marginal pricing (MTLMP)—and theoretically show that MDCP eliminates MWP (Proposition~\ref{prop:minDemand}) and MTLMP does so in most cases (Proposition~\ref{prop:MTLMPMW}). \tcb{By eliminating MWPs, these pricing schemes recover the cost of ramping-constrained dispatch through uniform in-market prices. }Moreover, the zero-MWP property of MDCP enables us to establish truthful bidding incentives for price-taking generators (Theorem~\ref{thm:bidRW}).

\textit{Empirical contributions:} First, we find that the relative efficiency of single- versus multi-interval dispatch depends fundamentally on the system's ramping capability and forecast error. \tcb{Multi-interval dispatch incurs higher upfront costs to secure ramping capacity but reduces penalties in scenarios with tight ramping requirements, whereas the advantage of multi-interval energy-ramp co-optimization over single-interval dispatch diminishes under large forecast errors and moderate ramping needs. Second, we observe that LMPs in rolling-window dispatch are highly volatile and frequently negative during constrained periods, causing under-compensation. Multi-interval dispatch yields higher prices during constrained ramp-down periods by reflecting opportunity costs associated with future intervals. Third, relative to LMP, the proposed uniform pricing mechanisms (MDCP and MTLMP) eliminate out-of-market MWPs, increasing generator profits  while raising total consumer payments. } Last, CAISO dataset exhibits higher net load volatility and more extreme ramping events than ERCOT, leading to larger differences in cost and pricing outcomes between dispatch and pricing methods, whereas ERCOT’s smoother net-demand profile results in smaller gaps and more stable pricing.

{\small
\begin{table}[htbp]
\vspace*{-9pt}
\caption{\small Major symbols}
\label{tab:symbols}
\vspace*{-10pt}
\begin{center}
\begin{tabular}{ll}
\hline
$N$ & total number of generators.\\
$T$ & length of the dispatch horizon.\\
$W$ & length of the rolling-window look-ahead horizon.\\
${\cal T}_{t'}$ & set of time intervals in the rolling window starting at $t'$.\\
${\cal G}_{t'}$ & energy--ramp co-optimization over ${\cal T}_{t'}$. defined by \eqref{eq:ED}.\\
$\hat{d}_t$ & forecast net demand at time $t$.\\
$\overline{\omega}_t,\underline{\omega}_t$ & upward and downward flexible ramping requirements.\\
$r_i^{\U},r_i^{\D}$ & physical ramp-up and ramp-down limits of generator $i$.\\
$g_{it},\overline{r}_{it},\underline{r}_{it}$ & generation, ramp-up capacity, and ramp-down capacity.\\
$c_{it},c_{it}^{\dagger}$ & bid-in marginal cost and true marginal cost of generator $i$.\\
${\cal M}_{it}(\cdot)$ & interval-based make-whole payment (MWP).\\
$J(\cdot)$ & penalty for energy \& ramping shortage, and   curtailment.\\
\hline
\end{tabular}
\end{center}
\vspace*{-10pt}
\end{table}
}

% \section{Pricing single-interval flexible  ramping}
% \input{SModel_v0}

%\vspace{-0.2cm}
\section{Energy and flexible ramping co-optimization}
\label{sec:A_Access}
In real-time operations, system operators forecast demand and issue dispatch instructions for energy and ramping products at five- or fifteen-minute intervals. Real-time operations vary significantly regarding look-ahead horizons. For instance, MISO employs a \textit{single-interval dispatch} based on a short-term forecast \cite{MISO:Schedule29}, whereas the CAISO utilizes a \textit{multi-interval dispatch} framework that optimizes trajectories over several future intervals \cite{CAISO:23FRP}. In this section, we review industry practices for energy and ramping procurement, detail the modeling assumptions adopted in this study, and present a general formulation for energy–ramping co-optimization that encompasses both single- and multi-interval dispatch.
 
 \vspace{-0.4cm}
 \subsection{\tcb{Background and Assumptions}}\label{sec:assumptions}

\tcb{We study rolling-widow real-time economic dispatch (RTED): in each interval, the operator solves a look-ahead dispatch over a window of future intervals, implements and settles only the binding interval, and re-solves in the next interval with updated forecasts and the realized dispatch as the initial condition. CAISO~\cite{CAISO:11FRP} and MISO~\cite{MISO:Schedule29} both co-optimize energy and ramping in RTED but differ in window length and in how they procure and price ramping capability. CAISO procures its FRP mainly in the 15-minute real-time unit commitment (RTUC) to secure maneuverability for its multi-interval 5-minute RTED, and settles it under a multi-settlement rule that allows financial buy-backs as real-time prices change. MISO instead procures ramp capability in its single-interval 5-minute energy and operating reserve market through a ramp capability demand curve, pricing it at its marginal opportunity cost so that the dispatch does not exhaust the ramp capability needed in upcoming intervals. Single-interval dispatch is a rolling window of one interval, so our setting covers both designs: we model CAISO's multi-interval window, with MISO's design as its special case.}

\tcb{Our goal is to analyze ramping procurement and BCR in RTED, not to replicate full-scale real-time operations. We therefore make five assumptions throughout. First, commitments are fixed: RTED runs after RTUC has fixed the commitment decisions~\cite{CAISO:22BCR}, and real-time prices are generally computed in RTED with commitments fixed, except under the fast-start pricing used by several ISOs. With fixed commitments and convex energy bids, RTED is a convex problem, so we focus on BCR caused by ramping constraints; uplift from unit-commitment nonconvexities, including startup and no-load costs, is outside our scope. Second, the network is a copper plate: the main text ignores transmission congestion and losses, and Sec.~VI-F of the online appendix~\cite{chen2026ramping} gives extensions to transmission-constrained networks. Third, ramping requirements are exogenous: they are set from net-demand forecasts (Sec.~\ref{sec:forecastmain}), and ramp capability demand curves are omitted. Fourth, settlement is simplified: cost-causation-based ramping settlements~\cite{CAISO:11FRP} are abstracted away, and generators receive energy and ramping payments directly from demand, which preserves the operator's revenue neutrality. Fifth, for the incentive analysis, suppliers are price-takers: each treats the shadow price of the power balance constraint as unaffected by its own offer. These assumptions simplify operations but preserve the  ramping limits, driving analysis for ramping procurement and BCR.}

\vspace*{-10 pt}%
\subsection{Energy-ramping co-optimization}
\begin{figure}[htbp]
    \centering
    \includegraphics[width=0.53\textwidth]{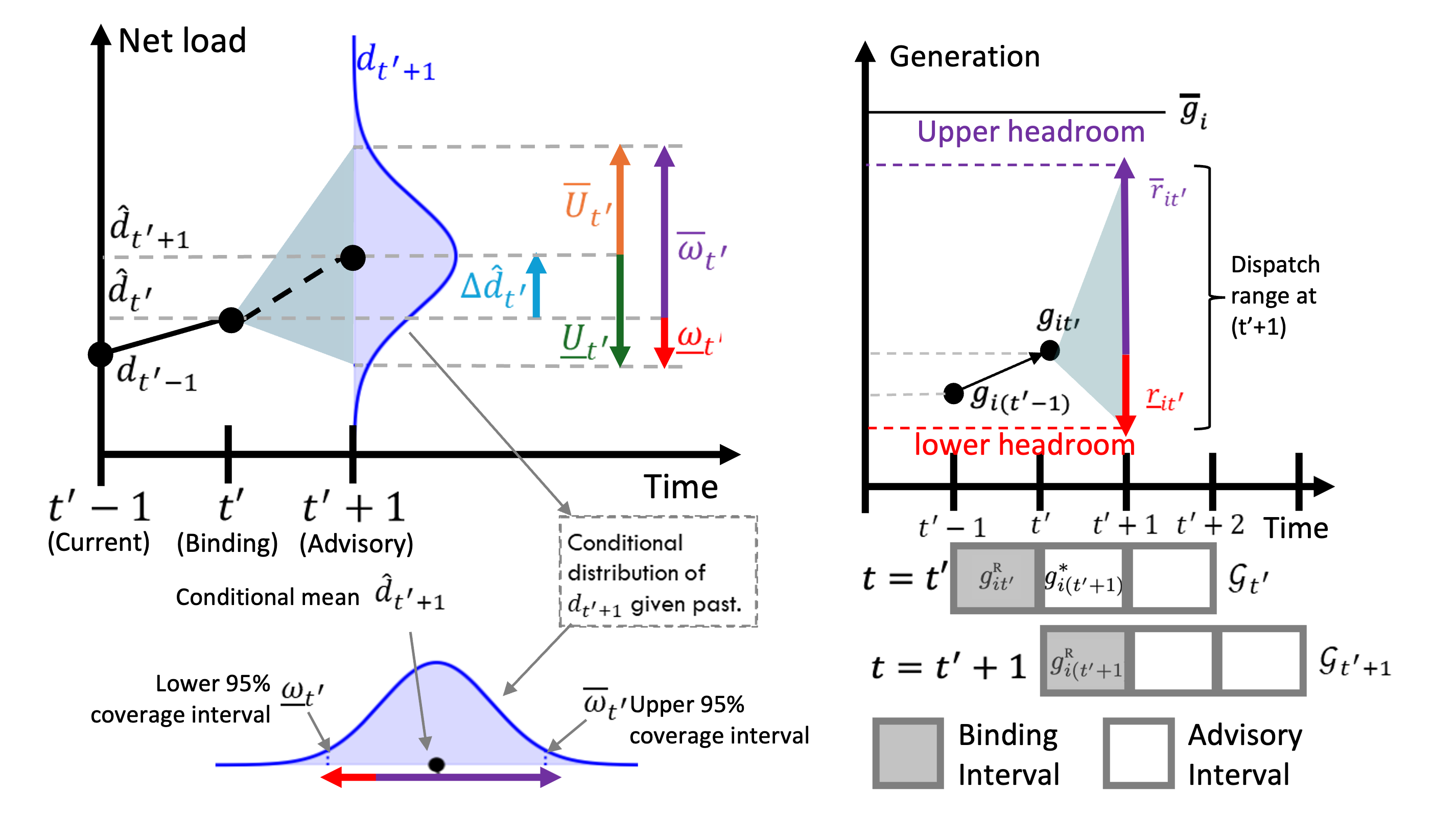}
    \vspace*{-20 pt}
    \caption{\scriptsize  Top left: Net demand trajectory with ramping product requirements over the look-ahead horizon ${\cal T}_{t'}$ and  $W=1$.  Top right: Partition of generation capacity into energy, ramp-up, and ramp-down procurements. The feasible dispatch range for the current operating point $g_{it}$ includes the ramp up and ramp down capacity procurements.  Bottom left: Distribution of net demand  and the system-wide upward and downward flexible ramping requirements covering the 95\% confidence interval of the conditional forecast error. Bottom right: sequential rolling-window dispatch with binding and advisory time intervals for $W=3$. ${\cal G}_{t'}$ and ${\cal G}_{t'+1}$ are computed according to \eqref{eq:ED}. }
    \vspace*{-10 pt}
    \label{fig:RWramp}
\end{figure}

Let the complete dispatch horizon be $[T]:=\{1,..., T\}$. Denote ${\cal T}_{t'} := \{t', t' + 1, \dots, t' + W - 1\}$ as a rolling-window look-ahead horizon with $t'$ as the binding interval and the rest are advisory intervals (Fig.~\ref{fig:RWramp} right). This means that, in each rolling-window, the operator optimizes over a look-ahead horizon of $W$ periods.  $W = 1$ represents  single-interval dispatch ${\cal T}_{t'} = \{t'\}$; $W > 1$ represents multi-interval dispatch.

%\tcr{Editable version of this figure is in "Fig.1\_Ramp\_Requirement.pptx" here in the overleaf document Fig folder.} }

%the initial generation level $g_{i(t-1)}$ can be merged it into capacity limits. 

Let $[N] := \{1, \dots, N\}$ denote the set of all generators.  The operator solves the  optimization ${\cal G}_{t'}$, which is assumed to have a unique optimal solution, to determine the generation  $\mathbf{g}_{t'}:=(g_{it})_{  i \in [N],t \in {\cal T}_{t'}}$, ramping up capacity $\mathbf{\overline{r}}_{t'}:=(\overline{r}_{it})_{  i \in [N],t \in {\cal T}_{t'}}$, and ramping down capacity $\mathbf{\underline{r}}_{t'}:=(\underline{r}_{it})_{  i \in [N],t \in {\cal T}_{t'}}$.
\begin{subequations} \label{eq:ED}
\begin{align}
{\cal G}_{t'}:~  &~~\underset{\substack{\{\gbf_{t'}, \overline{\rbf}_{t'}, \underline{\rbf}_{t'}, \lbf_{t'},\\ \sbf_{t'},  \underline{\sbf}_{t'},\overline{\sbf}_{t'} \geq \mathbf{0}\}}}{\rm minimize} &&   (\sum_{i, t} c_{it} g_{it}) + J(\lbf_{t'}, \sbf_{t'},  \underline{\sbf}_{t'},\overline{\sbf}_{t'})\label{eq:obj}\\
& {\rm subject~to} && \forall i \in [N], \forall t \in {\cal T}_{t'},\nn\\
&\lambda_{t}: && (\sum_{i\in [N]} g_{it})+ s_t= \hat{d}_{t} + l_t, \label{eq:PB}\\ 
&\overline{\eta}_t: && (\sum_{i\in [N]}\overline{r}_{it}) + \overline{s}_t\geq \overline{\omega}_t,  \label{eq:RBU}\\ 
&\underline{\eta}_t: &&  (\sum_{i\in [N]}\underline{r}_{it}) + \underline{s}_t \geq \underline{\omega}_t,  \label{eq:RBD}\\ 
& \tcb{t<t'+W-1:} && \tcb{-\underline{r}_{it}\le g_{i(t+1)}-g_{it} \le \overline{r}_{it}}\label{eq:ramp}\\%& & t' \le t \le t'+W-1,\\
&  && -r^{\D}_{i}\le g_{it'}-g_{i(t'-1)} \le r^{\U}_{i}, \label{eq:iniramp}\\
&  && \underline{r}_{it}\le g_{it}\le \overline{g}_i -\overline{r}_{it}, \label{eq:capa}\\
&   &&  0 \le \overline{r}_{it} \le r^{\U}_{i}, \quad 0 \le \underline{r}_{it} \le r^{\D}_{i}. \label{eq:Rcapa} 
\end{align}
\end{subequations}
 The objective \eqref{eq:obj} is to minimize the system operating cost. $c_{it}$ is the marginal generation cost from generator $i$'s offer at time $t$. \tcb{For simplification, we include linear generation costs; all theoretical results extend to the convex piecewise-affine costs widely used in practice~\cite{ChenTong25PESGM}.}   $l_t $ represents renewable curtailments,  and  $s_t$, $\overline{s}_t$, and $\underline{s}_t$  are slack variables associate  with energy balance and ramping requirements. These slack variables signal scarcity events; when they become nonzero, scarcity penalties are triggered in the objective through administratively determined penalty prices, including load shedding penalty $(p_t)$,  ramp down and up  shortfall penalties $(\overline{p}_t, \underline{p}_t)$, and renewable curtailment penalty $m_{t}$. These penalty prices are typically much higher than the highest generation cost $c_{it}$ with  $J(\lbf_{t'}, \sbf_{t'},  \underline{\sbf}_{t'},\overline{\sbf}_{t'}):=\sum_{t\in {\cal T}_{t'}}(p_{t}s_t+ \overline{p}_{t} \overline{s}_t+ \underline{p}_{t} \underline{s}_t+ m_{t} l_t) $.  
 
 Constraints \eqref{eq:PB} ensure supply–demand balance, while \eqref{eq:RBU}\eqref{eq:RBD} procures sufficient ramp-up and ramp-down capacities. \tcb{\eqref{eq:ramp} is the ramping constraint with ramping procurements for $t<t'+W-1$, since the last interval in the window has no successor.  \eqref{eq:iniramp} specifies the initial ramping constraint based on unit's ramping capability $r^{\U}_{i}$ and $r^{\D}_{i}$, since the dispatch at $t'-1$ is already fixed. Because all intervals have the same duration $\Delta t$, we set $\Delta t=1$ throughout, so the objective, constraints, and payment equations in the following sections are written in per-interval units.} Constraint \eqref{eq:capa} defines the feasible dispatch range for energy, which is smaller than the physical generation capacity $\overline{g}_i$ because upper and lower headrooms are reserved for ramping procurement (Fig.~\ref{fig:RWramp}, top right). The upper headroom represents reserved capability to increase generation, limited by both the unit’s maximum capacity $\overline{g}_i$ and its upward ramp rate. Symmetrically, the lower headroom ensures the generator can safely decrease its output without violating its minimum operating limits or downward ramp rates.  Constraint \eqref{eq:Rcapa} restricts procured ramping capacities according to the physical ramping limits   $r^{\U}_i$ and $r^{\D}_{i}$. The dual variables are denoted by $\lambda_t, \overline{\eta}_t$, and $\underline{\eta}_t$.%, $\underline{\gamma}_{it}$, and $\overline{\gamma}_{it}$. 

\vspace*{-10 pt}%
\subsection{Rolling-window dispatch}
\textcolor{blue}{Real-time dispatch is a sequential process with rolling-windows (Fig.~\ref{fig:RWramp} bottom right).  
 The operator is given the following forecasts and system parameters (Fig.~\ref{fig:RWramp} left): (i) net demand forecast $\{\hat{d}_t\}_{ t \in {\cal T}_{t'}}$ (defined as load minus variable renewable generation and assumed to be nonnegative), (ii) ramp up requirements $\{\overline{\omega}_t\}_{ t \in {\cal T}_{t'}}$ and  ramp down requirements $\{\underline{\omega}_t\}_{ t \in {\cal T}_{t'}}$, and (iii) the initial generation level $\{g_{i(t'-1)}\}_{i \in [N]}$.    The detailed forecast method are introduced in Sec.~\ref{sec:forecastmain}.
Then, the power system operator solves problem ${\cal G}_{t'}$ in \eqref {eq:ED}, and implements only the  {\em  binding} interval decisions:}
\beq\label{eq:binding}
g^{\R}_{it'} := g^*_{it'},~~ \underline{r}^{\R}_{it'} :=  \underline{r}^*_{it'},~~ \overline{r}^{\R}_{it'} := \overline{r}^*_{it'},
\eeq 
where superscript * denotes the optimal solution for energy dispatch and ramping procurement.  

For the next rolling window to obtain the dispatch at the binding interval $t'+1$,  the operator update the known parameter $g_{it'}=g^{\R}_{it'}$ for the new initial generation based on the solution \eqref{eq:binding} from the last rolling window. The operator updates the forecast $\{\hat{d}_{t}\}_{t \in {\cal T}_{t'+1}}$, ramp up requirements  $\{\overline{\omega}_t\}_{ t \in {\cal T}_{t'+1}}$, and ramp down requirements and $\{\underline{\omega}_t\}_{ t \in {\cal T}_{t'+1}}$ based on the real time information. Then, with  updated parameters, operator resolves ${\cal G}_{t'+1}$ via \eqref{eq:ED}.

For the  {\em single interval dispatch} with $W =1$, \eqref{eq:binding} represents the single-interval optimal dispatch decision in \eqref{eq:ED}.  For the {\em multi-interval dispatch} with $W >1$, although \eqref{eq:ED} produces optimal solutions over the entire horizon ${\cal T}_{t'}$, only the first {\em binding} interval results at $t'$ are implemented, while subsequent rolling-window decisions serve as {\em advisory} signals.  %Therefore, the rolling-window multi-interval dispatch signal is given by  \eqref{eq:binding}, which indicates that only partial solution is implemented and all future decisions are advisory. 

\textcolor{blue}{\vspace{-0.4cm}
\subsection{Net-Demand and Ramping Requirement Forecasting} \label{sec:forecastmain}
Before solving $\mathcal{G}_{t'}$ in \eqref{eq:ED}, the operator needs the net-demand forecast $\hat{d}_t$ and the upward and downward flexible ramping requirements $(\overline{\omega}_t,\underline{\omega}_t)$ for $t\in\mathcal{T}_{t'}$. Because $(\overline{\omega}_t,\underline{\omega}_t)$ cover the change from $t$ to $t+1$, these parameters depend on the conditional distribution of $(d_{t'},\dots,d_{t'+W-1})$ given past net demand. In this study, the predictor uses historical net demand observations only. We adopt a rolling-window linear minimum mean squared error (L-MMSE) predictor for net-demand forecast and using the standard forecasting errors to set the ramp requirements.  Specifically,
%We model net demand as a Gaussian process, for which the minimum mean squared error (MMSE) predictor is linear. This linear MMSE (L-MMSE) predictor is updated in each rolling window, it provides both the forecast trajectory and its probabilistic conditional distribution, from which the ramping requirements are set.
\subsubsection{L-MMSE forecasting}
$\xbf^{(t-1)}:=[d_{t-1},\dots,d_{t-L-1}]^\top$ collects the $L+1$ most recent net demands and $\ybf^{(t-1)}:=[d_{t},\dots,d_{t+W-1}]^\top$ represents the net demands to be forecast for the future $W$ intervals. With known means $\mubf_x^{(t-1)},\mubf_y^{(t-1)}$ and covariance matrices $\Sigmabf_{xx}^{(t-1)}$, $\Sigmabf_{yx}^{(t-1)}=(\Sigmabf_{xy}^{(t-1)})^\top$, and $\Sigmabf_{yy}^{(t-1)}$ estimated from historical data, the L-MMSE forecast and its prediction error covariance are
\begin{align}
\hat{\ybf} &:= [\hat{d}_{t},\dots,\hat{d}_{t+W-1}]^\top = \mubf_y+\Sigmabf_{yx}\Sigmabf_{xx}^{-1}(\xbf-\mubf_x), \label{eq:LMMSE}\\
\Sigmabf_{y|x} &:= \mbox{\sf Cov}(\ybf-\hat{\ybf}) = \Sigmabf_{yy}-\Sigmabf_{yx}\Sigmabf_{xx}^{-1}\Sigmabf_{xy}, \label{eq:LMMSEcov}
\end{align}
where superscript $^{(t-1)}$ is dropped for simplicity. The forecast \eqref{eq:LMMSE} is unbiased, and the covariance in \eqref{eq:LMMSEcov} implies that the predicted net-demand $\hat{d}_{t+k}$ has the same mean as $d_{t+k}$ but a smaller variance. 
\subsubsection{Ramping requirements}
The ramping requirements cover the forecast change from $t$ to $t+1$ plus the margins $(\overline{U}_t,\underline{U}_t)$ of a 95\% interval \cite{CAISO:25FRP_DMM} for $d_{t+1}$ (Fig.~\ref{fig:RWramp}, bottom left):
\begin{align}
&\Pr\big(\hat{d}_{t+1}-\underline{U}_t\le d_{t+1}\le\hat{d}_{t+1}+\overline{U}_t\,\big|\,\xbf^{(t-1)}\big)=0.95,\\
&\overline{\omega}_t=[\hat{d}_{t+1}-\hat{d}_t+\overline{U}_t]_+,\quad
\underline{\omega}_t=[\hat{d}_t-\hat{d}_{t+1}+\underline{U}_t]_+.
\end{align}
The general case for all ramp reserve requirements
$\{(\underline{\omega}_{t+k}, \overline{\omega}_{t+k})\}$ follows the same derivation. Hence $\overline{\omega}_t\ge\Delta\hat{d}_t:=\hat{d}_{t+1}-\hat{d}_t$, i.e., the upward requirement always covers the expected ramp (Fig.~\ref{fig:RWramp}, top left).  Note that the conditional variance of $d_{t+k}$ tends to increase with $k$, indicating increasing uncertainty as the forecast net demands moves further into the future. In practice~\cite{CAISO:23FRP}, however,
future ramp requirements may be set by the net-demand
forecasts (the conditional means) and the conditional variance $\sigma^2_{t|t-1}$ of the binding interval, which trades ramp violation risks with lowered ramp costs. These two ramping forecast settings are compared in Sec.~\ref{sec:model_comparison}

When $(d_t)$ is a Gaussian process, \eqref{eq:LMMSE}--\eqref{eq:LMMSEcov} give the exact conditional distribution $d_{t+k}\,|\,\xbf^{(t-1)}\sim\Nc(\hat{d}_{t+k},\sigma^2_{t+k|t-1})$, where conditional variance $\sigma^2_{t+k|t-1}$ is the $(k+1)$th diagonal entry of $\Sigmabf^{(t-1)}_{y|x}$.
If net demand is non-Gaussian, $\hat{d}_{t+k}$ remains unbiased but is generally not the conditional mean of $d_{t+k}$.  However, the actual conditional and unconditional variances of $d_{t+k}$ are smaller than $\sigma^2_{t+k|t-1}$. This may justify, weakly, the approximation of the actual conditional distribution of $d_{t+k}$ by $\Nc(\hat{d}_{t+k},\sigma^2_{t+k|t-1})$ and use the same ramp parameter setting procedure for the Gaussian case.  Note, however, because  $\sigma^2_{t+k|t-1}$ is likely larger than the actual conditional variance, the uncertainty interval is larger and the ramp reserve more conservative.    To obtain a theoretically accurate way to compute is through generative probabilistic forecasting, where a nonlinear mapping can be used to produce samples of the conditional distribution of $d_{t+k}$, from which the coverage interval can be computed through Monte Carlo methods  \cite{Tong&Wang:26NOW}. 

}

\section{\textcolor{blue}{Energy and flexible ramping prices}}
%\vspace{-0.2cm}
% \section{Uniform pricing avoiding make-whole uplift}\label{sec:AccessRight}
{\color{blue}To ensure that generators follow real-time dispatch signals, system operators combine in-market payments, determined by real-time prices such as the locational marginal price (LMP), with out-of-market payments (or uplifts). The latter are settled outside the dispatch optimization and compensate generators dispatched at prices below their bids.}%, a shortfall that arises under uncertainty and ramping constraints~\cite{CAISO:25BCRMIO}.}
\vspace*{-6 pt}%
\subsection{\textcolor{blue}{Principles of Real-Time Electricity Pricing}}\label{sec:principles}
{\color{blue}
Four principles, originating in classical market design and consistent with FERC's goals~\cite{FERC15upliftE-2_14}, guide our pricing design; other electricity markets may emphasize different principles.

\textit{(P1) Economic efficiency:} prices induce efficient production, consumption, and network use in the short run~\cite{Schweppe:88}.

\textit{(P2) Individual rationality:} no price-taking supplier loses by following the operator's dispatch.\footnote{\textcolor{blue}{Sec.~\ref{sec:OMP} defines metrics that quantify violations of (P2).}}

\textit{(P3) Incentive compatibility:} price-taking suppliers maximize their profits by bidding their true costs, so that efficient operation does not rely on altruism.\footnote{\textcolor{blue}{A supplier is price-taking if it treats the shadow price of the power balance constraint as unaffected by its own offer~\cite{Conejo:02TPS}.}}

\textit{(P4) Non-discrimination and transparency:} suppliers of products with the same features receive the same public price.

Wilson defines the system-wide real-time price as ``the highest price among those offers accepted in merit order to balance energy''~\cite{wilson2002architecture}. Without congestion and ramping limits, this price equals LMP (Sec.~\ref{sec.LMPRamp}) and satisfies (P1)--(P4).\footnote{\textcolor{blue}{The LMP reflects the marginal generation cost (P1); every accepted offer is paid at least its offer price (P2); a price-taker's bid does not affect the LMP and truthful-bidding is optimal (P3); and the LMP is uniform and public (P4).}}

Ramping limits and forecast errors break this equivalence: the operator may dispatch a supplier out of merit \cite{CAISO:25BCRMIO}, so the LMP can fall below its offer, violating (P2). Out-of-market uplift restores (P2) but compensates suppliers individually, violating (P4); because the uplift grows with each supplier's own offer, even a price-taking supplier can gain by inflating its offer, violating (P3). Combining the LMP with uplift thus satisfies (P1) and (P2) at the expense of (P3) and (P4).

This paper instead restores (P2) within the market through a uniform public price. In the California debate, Kahn et al.\ defended uniform pricing against pay-as-bid, its discriminatory alternative, arguing that pay-as-bid would alter bidding behavior in ways that harm efficiency~\cite{KAHN200170}. Hogan argues that energy-market deficiencies should be fixed first through real-time pricing, with out-of-market interventions as a second-best remedy~\cite{Hogan14EJ}. Following this route, we propose two uniform pricing methods, both coincide with LMP when ramping constraints are nonbinding. We keep Wilson's rule when ramping breaks merit order, setting the uniform price at the highest offer price among dispatched suppliers. %Hence (P2) holds without the out-of-market uplift of Sec.~\ref{sec:OMP}, (P4) is preserved, and Theorem~1 in Sec.~\ref{sec:TrueBid} establishes (P3) locally; MTLMP eliminates the uplift in most cases.
}

\vspace*{-7 pt}%
\subsection{In market payment}\label{sec.LMPRamp}
Real-time uniform pricing signals determine in-market payments for both energy and FRP. Following the marginal pricing principle, the real-time energy price LMP at the time interval $t$ derived from envelope theorem is given by
\beq\label{eq:LMPM}
\pi^{\text{\tiny LMP}}_{t} := \frac{\partial F}{\partial \hat{d}_{t}} = \lambda^*_{t},
\eeq
where $F$ denotes the optimal objective value in \eqref{eq:ED} and $ \lambda^*_{t}$ is the optimal dual associated with the balance constraint \eqref{eq:PB}.  

The up and down FRP prices $(\pi^{\U}_{t},\pi^{\D}_{t})$  are similarly obtained by envelope theorem, i.e.,
\beq\label{eq:rampprice}
\pi^{\U}_{t}:= \frac{\partial F}{\partial \overline{\omega}_{t}}=\overline{\eta}^*_t,~~ \pi^{\D}_{t}:= \frac{\partial F}{\partial \underline{\omega}_{t}}=\underline{\eta}^*_t,
\eeq
following principles analogous to reserve capacity pricing \cite{CAISO:11FRP, WuPapalexopoulos04TPSreservepricing}. $\overline{\eta}^*_t$ and $\underline{\eta}^*_t$  are the optimal dual variables of the upward and downward ramping constraints \eqref{eq:RBD}, respectively.
  
The total in-market payment to generator $i$ at the time interval $t$ consists of an energy payment ${\cal P}^{\text{\tiny LMP}}_{it} $ and a ramp capability payment ${\cal R}_{it}$, \ie
    \beq\label{eq:energypay}
    {\cal P}^{\text{\tiny LMP}}_{it} := \pi^{\text{\tiny LMP}}_{t} g_{it}^{\R},
    \eeq
    \beq
    {\cal R}_{it} := \pi^{\U}_{t} \overline{r}^{\R}_{it}  + \pi^{\D}_{t} \underline{r}^{\R}_{it} \label{eq:ramppay},
    \eeq
where $g_{it}^{\R}$, $\overline{r}^{\R}_{it} $, and $\underline{r}^{\R}_{it} $ denote the {\em binding} real-time dispatch of energy, upward ramping, and downward ramping, respectively, as determined by \eqref{eq:binding}.

%In practice, real-time market settlements for energy and ramping are more complex, accounting for metering granularity, incentive compatibility, ramping cost causation, and multi-settlements arising from forecast updates \cite{CAISO:11FRP, FERC15upliftE-2_14}. 

\vspace*{-8 pt}%
\subsection{Out of market payment}\label{sec:OMP}
\tcb{Due to uncertainty and ramping  coupling, LMP alone may fail to provide sufficient dispatch-following incentives for all generators. Out-of-market payments—namely make-whole payments (MWP) and lost opportunity costs (LOC)—address this issue.} MWP compensates generators for underpayment. LOC compensates for foregone profits from alternative optimal dispatches. LOC provides stronger incentive alignment, while MWP serves as a baseline measure ensuring immediate compensation. In practice, MWP is used more frequently by assuming generators with immediate compensation are willing to follow the real-time dispatch signal. 

%By paying interval-based MWP to generators, we directly eliminate under compensation at each time interval to all generators. MWP establishes the immediate dispatch-following incentives---generators observe the binding interval price and make decisions to avoid immediate under compensation. We assume generators are willing to follow the dispatch signal if the price is no less than it's bids, which is equivalent to MWP=0. 

Denote ${\cal M}_{it}(\cdot)$ as the interval-based MWP\footnote{In practice, daily MWPs are often adopted for real-time BCR uplifts, \ie $\max\{0, \sum_{t \in {\cal H}} (c_{it} g^{\R}_{it})- \pi_t g^{\R}_{it}\}$ \cite{CAISO:24BCR}. We compute both interval-based MWPs and daily MWPs in the simulation and  the online appendix  \cite{chen2026ramping}. } for generator $i$ at time $t$, which can be computed by  
 %Ignoring the price forecast ability, MWP directly removes under compensation at each time interval by
\beq\label{eq:mwp}
{\cal M}_{it}(\pi_t, g^{\R}_{it}) := \max\{0,  c_{i} g^{\R}_{it}- \pi_t g^{\R}_{it}\},
%{\rm MWP}_{it}(\pmb{\pi},\mathbf{g}^{\RED}_i) = \max\{0, \sum_{t \in {\cal H}} (\pi_t g^{\RED}_{it}-c_{it} g^{\RED}_{it})\}.
\eeq
where $\pi_t$ is the energy price. MWP is positive only when the price is less than it's bids ($\pi_t < c_{i}$). Thus, by paying MWPs, the operator eliminates under-compensation. Note that out of market payment MWP is discriminative over different generators, although LMP for in market payment is uniform.% which makes the energy-ramping reserve payments discriminative. We say a uniform price $\pi_t$ supports dispatch-following incentives if $ {\cal M}_{it}(\pi_t, g^{\R}_{it})=0, \forall i  \in [N], \forall t\in[T]$. 

\tcb{Definition \eqref{eq:mwp} counts only energy revenue: it tests whether the energy price alone covers each dispatched unit's energy bid cost, which is what the pricing schemes in this paper are designed to achieve. Ramping payments \eqref{eq:ramppay} compensate a separate product at its opportunity cost, and counting them would let a ramping payment mask an energy-price shortfall. Settlement designs differ on this point, but because ramping payments are nonnegative, \eqref{eq:mwp} upper-bounds the MWP with ramping revenue included, under any pricing scheme. Note that, under rolling-window LMP, ramping revenue reduces but need not eliminate the MWP (Appendix~VI-E\footnote{\tcb{In the example of Table~XI, G1 receives a ramping payment of 4{,}000 under M-LMP but earns $420 \times (10 - 25) = -6{,}300$ in the energy market at the binding interval $t = 1$, leaving a deficit of 2{,}300. Hence MWP/BCR can remain necessary under LMP even when ramping revenues are counted. }} of~\cite{chen2026ramping}).} %Only the least-cost property of Proposition~\ref{prop:minDemand} relies on the energy-only definition, since counting ramping revenue loosens the zero-MWP constraint and may admit a uniform price below MDCP.

Eliminating out-of-market payments improves market transparency, as such payments often distort real-time uniform price signals and incentivize strategic bidding \cite{CAISO:25BCRMIO}. As shown in \cite{Guo&Chen&Tong:21TPS}, no uniform price can remove LOC for all generators. Nonetheless, uniform prices are preferred for transparency, and we next propose two such mechanisms—MDCP and MTLMP—that achieve zero MWP.

%\tcr{MISO single interval total energy and ramping payments are proved to be higher than the generation cost. Shall we say MISO also has under-compensation for just energy payment part?}
 
% \subsubsection{MWP under LMP+FRP}
% LMP+FRP fails to support MWP=0 when there’re more than 1 generators with binding ramping constraints.

\vspace*{-10 pt}%
\subsection{Max Dispatch Cost Pricing (MDCP)}

Define the real-time MDCP at time $t$ as
\beq\label{eq:mdcp}
\pi^{\mDP}_{t} :=  {\rm max} ~~ \{c_{1t}\mathbbm{1}_{[g^{\R}_{1t}>0]}, ..., c_{Nt}\mathbbm{1}_{[g^{\R}_{Nt}>0]}, p_{t}\mathbbm{1}_{[s^{\R}_{t}>0]}\},
\eeq
where $\mathbbm{1}_{[g^{\R}_{it}>0]}$ indicates the dispatch status of generator $i$. The indicator   $\mathbbm{1}_{\cal X}$ equals 1 if ${\cal X}$ is true, otherwise it's zero.
In normal operation, MDCP  equals the maximum marginal cost among dispatched generators; in scarcity events with nonzero slack variable $s^{\R}_{t}$ for load shedding, MDCP equals to the load shedding penalty prices $p_t$. The superscript $^{\R}$, consistent with  \eqref{eq:binding}, denotes the dispatch outcome realized in the binding interval of the rolling-window dispatch. Assume realized demand $d_t \in \mathbb{R}^+$, under MDCP, the MWP computed by \eqref{eq:mwp} are eliminated for all generators. \tcb{Among all uniform prices that yield zero MWP for every generator and are no lower than the penalty whenever load is shed, MDCP minimizes the  demand payment. This result is  stated below.}
 
%As illustrated by Proposition~\ref{prop:minDemand}, we derive a closed-form solution for MDCP in \eqref{eq:mDPdemand}. 
 \begin{proposition}\label{prop:minDemand}
\tcb{MDCP yields zero MWP for every generator and minimizes the total demand payment among all uniform prices $\pi_t$ that satisfy ${\cal M}_{it}(\pi_t, g^{\R}_{it})=0$ for all $i \in [N]$ and $t \in [T]$, and $\pi_t \geq p_t \mathbbm{1}_{[s^{\R}_{t}>0]}$ for all $t \in [T]$.}
\end{proposition} 
 
 The proof is in the online appendix  \cite{chen2026ramping}. \tcb{Here, the real-time dispatch $g^{\R}_{it}$ is computed by \eqref{eq:binding}, independently of the pricing rule. MDCP therefore changes prices but not the dispatch, preserving economic efficiency, the first pricing principle (P1) in Sec.~\ref{sec:principles}. Under MDCP, generator $i$ receives at time $t$ the in-market energy payment ${\cal P}^{\text{\tiny MDCP}}_{it} := \pi^{\text{\tiny MDCP}}_{t} g_{it}^{\R}$ and the ramping payment \eqref{eq:ramppay}.  Proposition~\ref{prop:minDemand} shows that no out-of-market payment is needed: MDCP eliminates BCR through a uniform public price, satisfying individual rationality (P2) and non-discrimination and transparency (P4), at the lowest demand payment among uniform prices. Sec.~\ref{sec:TrueBid} further shows that MDCP satisfies incentive compatibility (P3) locally.}
%Note that, the generalization of MDCP to network-constrained cases is discussed in the appendix. MDCP cannot be directly extended to this case, but our network extension of MDCP still minimizes demand payment with nodal uniform pricing while removing MWP. %In the next section, we propose MTLMP with a natural network extension.

%Reason: ramp requirement confidence interval is larger than demand forecast (mean). If binding ramping for ramping requirement, then nonbinding ramping for the demand forecast. (rewrite the demand match in CAISO future intervals as a ramp procurement constraint.

%for ramp up case s-LMP = MDCP

%\textbf{Conjecture:} In ramp up case, MDCP$\le $ LMP in the single interval case (MISO Case). Both MDCP and LMP has zero MWP. 

%LMP$>$MDCP when using ramp reserve at the reserve price.
%\tcr{Do I need to show MTLMP = MDCP at some places?}
%Therefore, among all uniform pricing avoiding MWP, MDCP actives the minimum demand payment.  

% demand payment under LMP + single interval MWP can be less than the demand payment under this uniform price, although this uniform price won’t have MWP. This uniform price has the least demand payment among all uniform prices that won’t have MWP. 
%\tcr{Austran Cal Mounger... marginal cost production pricing}

%\vspace{-0.3cm}

\vspace*{-10 pt}%
\subsection{Max Temporal Locational Marginal Pricing (MTLMP)}

%  \beq \label{eq:mtlmp}
%  \begin{array}{l}
%  \pi^{\mTLMP}_{t}:= \underset{i \in {\cal N}}{\rm max}~~\pi_{it}^{\TLMP}\\
% ~~~~~~=  \lambda_{t}^*+\underset{i\in {\cal N}}{\rm max}  (-\underline  \gamma_{i(t+1)} ^*+\overline \gamma_{i(t+1)}^* +\underline \gamma_{it}^* - \overline \gamma_{it}^*\\
% ~~~~~~+\underline \mu_{i}^* - \overline \mu_{i}^*),
% \end{array}
%  \eeq

%  The KKT conditions from
% \eqref{eq:ED} gives
% \beq
% \begin{array}{lrl}\label{eq:EDKKT}
% && c_{it}-\lambda^*+\underline  \gamma_{i(t+1)} ^*-\overline \gamma_{i(t+1)}^* -\underline \gamma_{it}^* + \overline \gamma_{it}^*-\underline \mu_{i}^* + \overline \mu_{i}^*=0.
%  \end{array}
%  \eeq

Define the real-time MTLMP at time $t$ as
 \beq \label{eq:mtlmp}
 \begin{array}{l}
 \pi^{\mTLMP}_{t}:= {\rm max} ~~ \{  \pi_{1t}^{\TLMP},..., \pi_{Nt}^{\TLMP}, p_{t}\mathbbm{1}_{[s^{\R}_{t}>0]}\}\\
%~~~~~~~=  \lambda_{t}^*+\underset{i\in [N]}{\rm max}  (-\underline  \gamma_{i(t+1)} ^*+\overline \gamma_{i(t+1)}^* +\underline \gamma_{it}^* - \overline \gamma_{it}^*),
\end{array}
 \eeq
 where temporal locational marginal pricing\footnote{TLMP definition here is different but consistent with \cite{Guo&Chen&Tong:21TPS}, which computes the derivative at the optimal solution.} (TLMP)  is given by $\pi^{\TLMP}_{it}:= - \frac{\partial}{\partial g_{it}} V_{it}(g^*_{it}, \rbf^*)$ with $\mathbf{r}^*:=\{\mathbf{\overline{r}}^*_t, \mathbf{\underline{r}}^*_t\}_{t\in {\cal T}_{t'}}$ and $V_{it}(g^*_{it}, \rbf^*):=F(g^*_{it}, \rbf^*)-c_{it}g^*_{it}$. $\mathbf{r}^*$ represents the optimal ramping procurement. $V_{it}(g^*_{it}, \rbf^*)$ represents the optimal total generation cost in \eqref{eq:obj} excluding the contribution from generator $i$ in interval $t$. When deriving the energy price MTLMP, we fix all ramping products at the optimal solution to exclude the influence of ramping products. That way, in normal cases without load shedding, TLMP  evaluates the marginal energy cost of losing generation for generator $i$ at time $t$. Thus, MTLMP represents the system’s worst marginal cost of losing generation.  
In scarcity cases with nonzero slack variable $s^{\R}_{t}$ for load shedding, we set MTLMP equals the load shedding penalty $p_t$. Through the envelope theorem,  MTLMP can be computed by the optimal dual values of \eqref{eq:ED}. \tcb{The derivation of MTLMP, including the case in which \eqref{eq:ED} is degenerate and its dual solution is not unique, is in the online appendix~\cite{chen2026ramping}.}

\tcb{MTLMP have several advantages: (i) generators whose ramping constraints do not bind usually receive prices above their bid-in costs;} (ii) it  reflects the system’s marginal generation loss; (iii) it extends naturally to network-constrained systems; and (iv) it generally eliminates MWP (Proposition~\ref{prop:MTLMPMW}).

 \begin{proposition}\label{prop:MTLMPMW} Under MTLMP, for any generator $i$ at time $t$, if the dispatch does not equal its provided ramp-down capacity \ie $g^*_{it} \neq \underline{r}^{*}_{it}$, then ${\cal M}_{it}(\pi^{\mTLMP}_{t}, g^{\R}_{it})=0$. 
   \end{proposition}

The proof is in the online appendix  \cite{chen2026ramping}. \tcb{Under MTLMP, generator $i$ receives at time $t$ the in-market energy payment ${\cal P}^{\text{\tiny MTLMP}}_{it} := \pi^{\text{\tiny MTLMP}}_{t} g_{it}^{\R}$ and the ramping payment \eqref{eq:ramppay}. As with MDCP, the dispatch is unchanged, preserving economic efficiency (P1), and the price is uniform and public (P4). By Proposition~\ref{prop:MTLMPMW}, MTLMP also satisfies individual rationality (P2) except under pure ramp-down scheduling, i.e., when $g^*_{it}=\underline{r}^{*}_{it}$.  We do not establish incentive compatibility (P3) for MTLMP, because generators scheduled purely for ramp-down may still require out-of-market MWP, which can reward inflated offers. In our simulations (Table~\ref{tab:mwp_comparison}), MWP under MTLMP is 0.0\% in every scenario except S1, an extreme ramp-scarcity stress test, where MWP is 0.1--0.2\% of generators' total energy revenue.}

%\tcb{Under MTLMP, generator $i$ receives at time $t$ the in-market energy payment ${\cal P}^{\text{\tiny MTLMP}}_{it} := \pi^{\text{\tiny MTLMP}}_{t} g_{it}^{\R}$ and the ramping payment \eqref{eq:ramppay}. As with MDCP, the dispatch is unchanged, preserving economic efficiency (P1), and the price is uniform and public (P4). By Proposition~\ref{prop:MTLMPMW}, MTLMP also satisfies individual rationality (P2) except under  pure ramp-down scheduling $g^*_{it} \neq \underline{r}^{*}_{it}$. This condition is rare: in our simulations (Table~\ref{tab:mwp_comparison}), MWP under MTLMP is at most 0.1--0.2\% of generators' total energy revenue, in the most ramp-constrained case, and zero in nearly all other cases. We do not analyze incentive compatibility (P3) for MTLMP, whose temporal structure complicates the analysis of bidding incentives, and leave it for future work.}

\vspace{-0.3cm}
\subsection{Truthful Bidding Incentives for Price-Takers}\label{sec:TrueBid}

\tcb{Because MDCP is computed from bid-in costs, generators might gain by misreporting, e.g., by inflating their offers, violating incentive compatibility, the third pricing principle (P3) in Sec.~\ref{sec:principles}. The same question arises for classical single-interval LMP, under which bidding the true marginal cost is optimal for price-taking generators. Out-of-market uplift breaks this property even for price-takers, because the uplift depends on each generator's own offer. Numerical examples show generators inflating their offers to collect larger uplift under rolling-window LMP (Table~V in the appendix of~\cite{Guo&Chen&Tong:21TPS}; Table~II of~\cite{norambuenaguzman2026optimal}), and market data indicate suspected BCR exploitation~\cite{CAISO:22BCR}. In 2013, FERC approved a settlement resolving allegations that JP Morgan Ventures Energy Corporation used manipulative bidding strategies in CAISO and MISO to obtain above-market MWP~\cite{FERCJPM}; under one strategy, MWP/BCR made up about 53\% of what its units received from CAISO.\footnote{\textcolor{blue}{The strategy exploited ramp-rate constraints between consecutive day-ahead clearings to obtain ramping awards paid as bid. While it was in use (April 1 to June 18, 2011), the units received approximately \$26 million in BCR, \$17 million at market rates, and \$6 million from exceptional dispatches~\cite{FERCJPM}, so BCR accounted for $26/49 \approx 53\%$ of these payments.}} Because uplift is ultimately charged to load, such bidding raises demand payments. Under truthful bidding, LMP with BCR can yield a lower demand payment than MDCP, since BCR tops up each generator only to its own bid-in cost while uniform MDCP pays all generators the same higher price. However, strategic bidding can erase this advantage. Denote $\cbf^\dagger_{i}:=(c^\dagger_{it})_{t\in {\cal T}_{t'}}$ as generator $i$'s true marginal cost, and $\boldsymbol{\pi}^{\mDP}_{-i}:=(\pi^{\mDP}_{-it})_{t\in {\cal T}_{t'}}$ the MDCP   when generator $i$ is excluded from \eqref{eq:ED}. MDCP removes this channel by requiring no uplift, and Theorem~\ref{thm:bidRW} shows that, locally, the MDCP price itself does not reward misreporting.}

\begin{theorem}\label{thm:bidRW} Consider the one-shot problem  \eqref{eq:ED} over the rolling-window ${\cal T}_{t'}$. Suppose that (i) all generators are price-taking and profit-maximizing; (ii) the optimal dispatch is unique; and (iii) all generators other than generator $i$ bid truthfully.
Then truthful bidding is locally profit-maximizing for generator $i$ \ie  $\cbf_i=\cbf_i^\dagger$ maximizes profit of generator $i$ over the set ${\cal C}_i:=\{\cbf: \cbf^\dagger_{i} \le \cbf \le \overline{\cbf}_i\}, \overline{\cbf}_i:=\max\{\cbf^\dagger_{i},\boldsymbol{\pi}^{\mDP}_{-i} \}$.
\end{theorem}

\tcb{The proof is given in the online appendix~\cite{chen2026ramping}. Its key step shows that the MDCP markup over generator $i$'s marginal cost is the same for all bids in ${\cal C}_i$; by Topkis' monotonicity theorem~\cite{Topkis78}, raising the bid then weakly decreases the dispatched quantity and hence the profit. Because the theorem holds for every generator when all others bid truthfully, the truthful profile $(\cbf_1^\dagger,\dots,\cbf_N^\dagger)$ is a Nash equilibrium of the restricted game in which each generator $i$ chooses its bid from ${\cal C}_i$; we use ``local'' in this sense.}

\tcb{Theorem~\ref{thm:bidRW} considers upward deviations because MDCP is set by bid-in costs, so bid inflation is the main incentive concern. Sufficiently small downward deviations do not raise profit either: they leave the dispatch unchanged, and in each interval MDCP is either unchanged or, if generator $i$ sets it, falls below its true cost. Truthful bidding is therefore weakly optimal against small deviations in both directions in every interval where generator $i$ is inframarginal, i.e., $c^\dagger_{it}<\pi^{\mDP}_{-it}$. No such guarantee holds where it sets MDCP. The set ${\cal C}_i$ reduces to $\{\cbf_i^\dagger\}$ if and only if $\pi^{\mDP}_{-it}\le c^\dagger_{it}$ for all $t$. Then either generator $i$ is not dispatched and earns zero profit, because cheaper resources serve the demand, or it is dispatched and sets MDCP. In the latter case, raising its bid slightly above its true cost leaves its dispatch unchanged\footnote{\textcolor{blue}{Because \eqref{eq:ED} is a linear program with a unique minimizer, sufficiently small changes in its cost coefficients do not change the minimizer.}} but raises MDCP, strictly increasing its profit. The theorem thus holds trivially in the singleton case, and ${\cal C}_i$ cannot be enlarged in general.}

\tcb{Theorem~\ref{thm:bidRW} considers the one-shot dispatch \eqref{eq:ED}, in which the forecast for the whole window is known and generator~$i$ bids once. To extend its conclusion to rolling-window operation, we view bidding as generator~$i$'s sequential decision problem. Let $Q_i(\cbf_i;\boldsymbol{\xi})$ denote generator~$i$'s profit from bids $\cbf_i$ when the forecasts realize as $\boldsymbol{\xi}$. A clairvoyant generator that knew $\boldsymbol{\xi}$ in advance could attain $Q_i^*(\boldsymbol{\xi}):=\max_{\cbf_i\in{\cal C}_i} Q_i(\cbf_i;\boldsymbol{\xi})$. A rolling-window bidder instead updates its bids window by window using only the forecasts available at each window, so its bids $\cbf_i^{\R}(\boldsymbol{\xi})$ are non-anticipative. If these bids lie in ${\cal C}_i$, they are feasible for the clairvoyant problem, so $Q_i(\cbf_i^{\R}(\boldsymbol{\xi});\boldsymbol{\xi})\le Q_i^*(\boldsymbol{\xi})$ for every realization $\boldsymbol{\xi}$; this is the perfect-information relaxation of stochastic dynamic programming~\cite{Brown2010}, obtained by removing the non-anticipativity constraint. Now suppose that at some realization $\boldsymbol{\xi}^*$, generator~$i$'s profit under window-by-window clearing equals its profit in the one-shot problem for every bid in ${\cal C}_i$. Theorem~\ref{thm:bidRW} then gives $Q_i^*(\boldsymbol{\xi}^*)=Q_i(\cbf_i^\dagger;\boldsymbol{\xi}^*)$, so at $\boldsymbol{\xi}^*$ no bidding policy with bids in ${\cal C}_i$ earns more than truthful bidding. Hence no such policy dominates truthful bidding: a price-taking generator that bids without foresight of demand has no deviation in ${\cal C}_i$ that raises its profit at every realization. The argument does not rule out policies that raise expected profit by exploiting knowledge of the forecast-error distribution, which we leave for future work.}

\tcb{Theorem~\ref{thm:bidRW} has two limitations. First, it assumes price-taking, as does the classical incentive-compatibility result for single-interval LMP; strategic bidding by generators with market power is beyond the scope of this paper. Second, it is local, reflecting MDCP's dependence on dispatched offers: as shown above, a generator that sets MDCP can gain by bidding slightly above its true cost.}

\vspace{-0.2cm}
\section{Numerical Results}\label{sec:SOAccessRight}
%\vspace{-0.0015cm}
We compare single interval and rolling-window multi-interval dispatch frameworks under LMP, MDCP, and MTLMP. The performance is evaluated using metrics including MWP out of market payment, demand payment, and operation efficiency. %In the  appendix of the  supplemental material, we provide additional  results and toy examples for more  insights.% of ramping procurement and MWP.

\vspace{-0.3cm}
\subsection{Parameter Settings}

\begin{figure}[!htbp]
    \centering
    \vspace*{-9pt}%
    
    % --- Top plots: Side-by-side ---
    \includegraphics[width=0.24\textwidth]{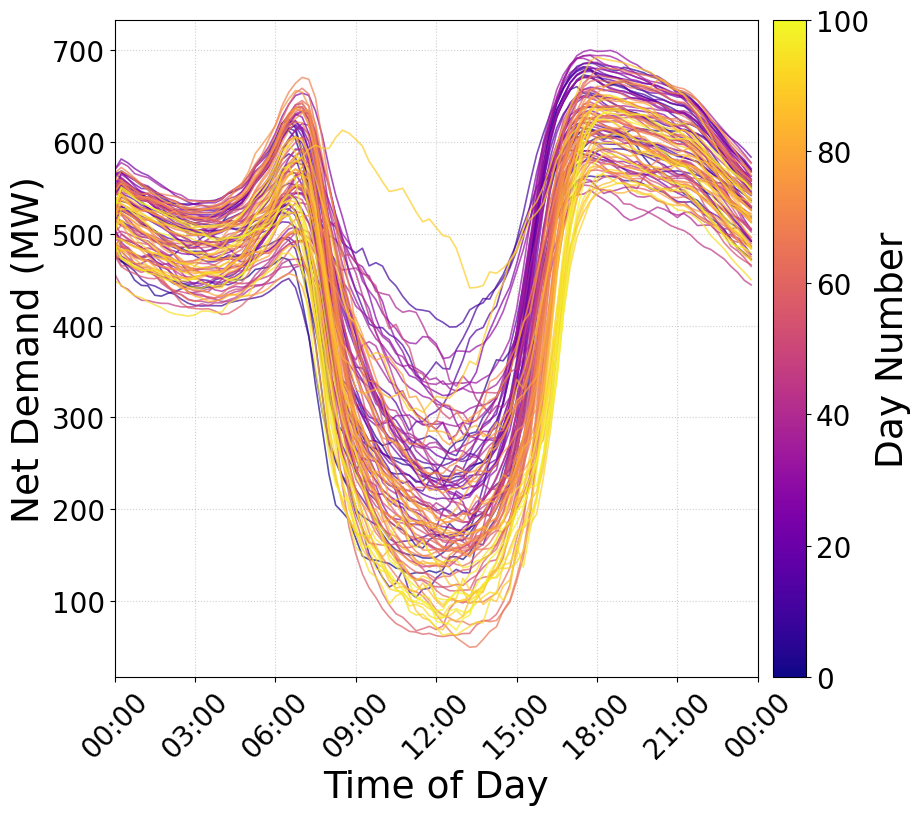}
    \hfill
    \includegraphics[width=0.24\textwidth]{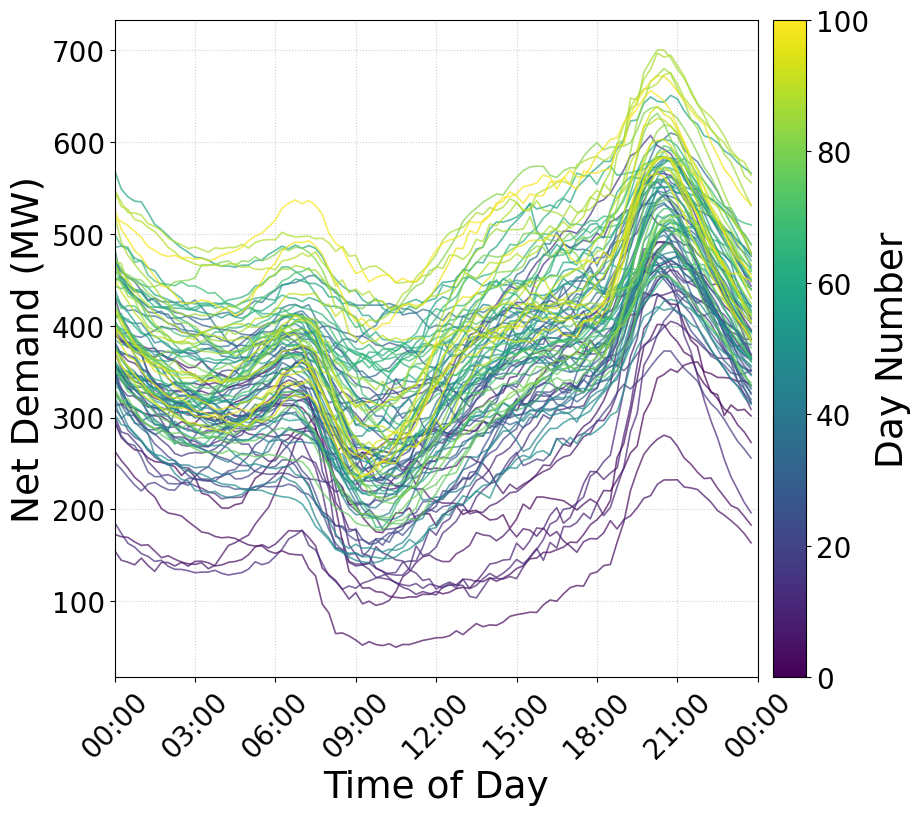}
    
    % --- Table below the plots ---
    \vspace*{10pt}%
    \begin{minipage}{0.49\textwidth}
        \centering
        \scalebox{0.8}{ % <--- scale factor (try 0.7–0.85)
        \begin{tabular}{|c|cccccccccc|}
        \hline
        \textbf{Ramp rate} & \textbf{S1} & \textbf{S2} & \textbf{S3} & \textbf{S4} & \textbf{S5} & \textbf{S6} & \textbf{S7} & \textbf{S8} & \textbf{S9} & \textbf{S10} \\ \hline
        \textbf{G1} & 20  & 30  & 50  & 50  & 50  & 50  & 50  & 75   & 100 & 499.9 \\ 
        \textbf{G2} & 15  & 15  & 15  & 30  & 50  & 50  & 50  & 75   & 100 & 499.9 \\ 
        \textbf{G3} & 15  & 15  & 15  & 15  & 15  & 30  & 50  & 75   & 100 & 499.9 \\ \hline
        \end{tabular}
        }
    \end{minipage}
    
    \caption{\scriptsize Top Left: Realized net-demand trajectories for the 100-day simulation period from CAISO data. Top Right: Realized net-demand trajectories from ERCOT data. Bottom: Ramping scenario settings. Unit is MW/15-minute. S1-S10 represent different ramping scenarios. G1-G3 represent different generators.}
    \label{fig:param}
    \vspace*{-9 pt}%
\end{figure}

\tcb{We simulated 100 days of 15-minute net-demand trajectories from CAISO~\cite{CAISOData} and ERCOT (system load minus wind and solar generation~\cite{ERCOTLoadData,ERCOTWindData,ERCOTSolarData}), starting November 15, 2024, and May 1, 2025, respectively (Fig.~\ref{fig:param}, top). Both datasets are affinely rescaled to a common net-demand range.\footnote{Specifically, we re-scaled the data through an affine transformation such that both datasets share the exact same minimum and maximum values throughout the entire 100-day period.}    The critical periods for flexibility in CAISO are the steep morning ramp-down and the sunset ramp-up; ERCOT trajectories are flatter, with variability driven mainly by wind uncertainty.}

\tcb{In each rolling window ${\cal T}_{t'}$, the look-ahead dispatch \eqref{eq:ED} determined the dispatch and the upward and downward ramp capability of the binding interval $t'$. Each interval lasted 15 minutes, so per-interval quantities were converted to energy and payments with $\Delta t=0.25$~h. The multi-interval dispatch used $W=4$, i.e., a one-hour window~\cite{CAISO:23FRP}, with ramp capability procured for every interval in the window. Net-demand forecasts and ramping requirements followed Sec.~\ref{sec:forecastmain} with $L+1=96$. We implemented the models in Pyomo~6.9.0 and solved them with Gurobi~12.0.1 using default numerical tolerances, on a 64-bit Qualcomm ARMv8 processor with 16~GB of RAM. More implementation details are given in Sec.\ref{sec:AddSim} of the online appendix~\cite{chen2026ramping}.}%We compared the two margin settings defined there: M, whose margins widen over the window, and 15-m, which applies the binding-interval margins to all advisory intervals while retaining each interval's expected net-demand change. 

\tcb{We considered a copper-plate system with three generators, G1--G3, with bid-in marginal costs of 25, 30, and 50~\$/MWh and capacities $\overline{g}_i=500$~MW. The penalty costs for ramping shortfalls and curtailment were 80~\$/MWh~\cite{CAISO:20FRPRefinements}. Generator ramping capabilities were varied across ten configurations, S1--S10, from lowest to highest (Fig.~\ref{fig:param}, bottom).}

\vspace{-0.3cm}

\subsection{Price comparison}

\begin{figure}[htbp]
    \centering
    \vspace{-0.3cm}
    \includegraphics[width=0.5\textwidth]{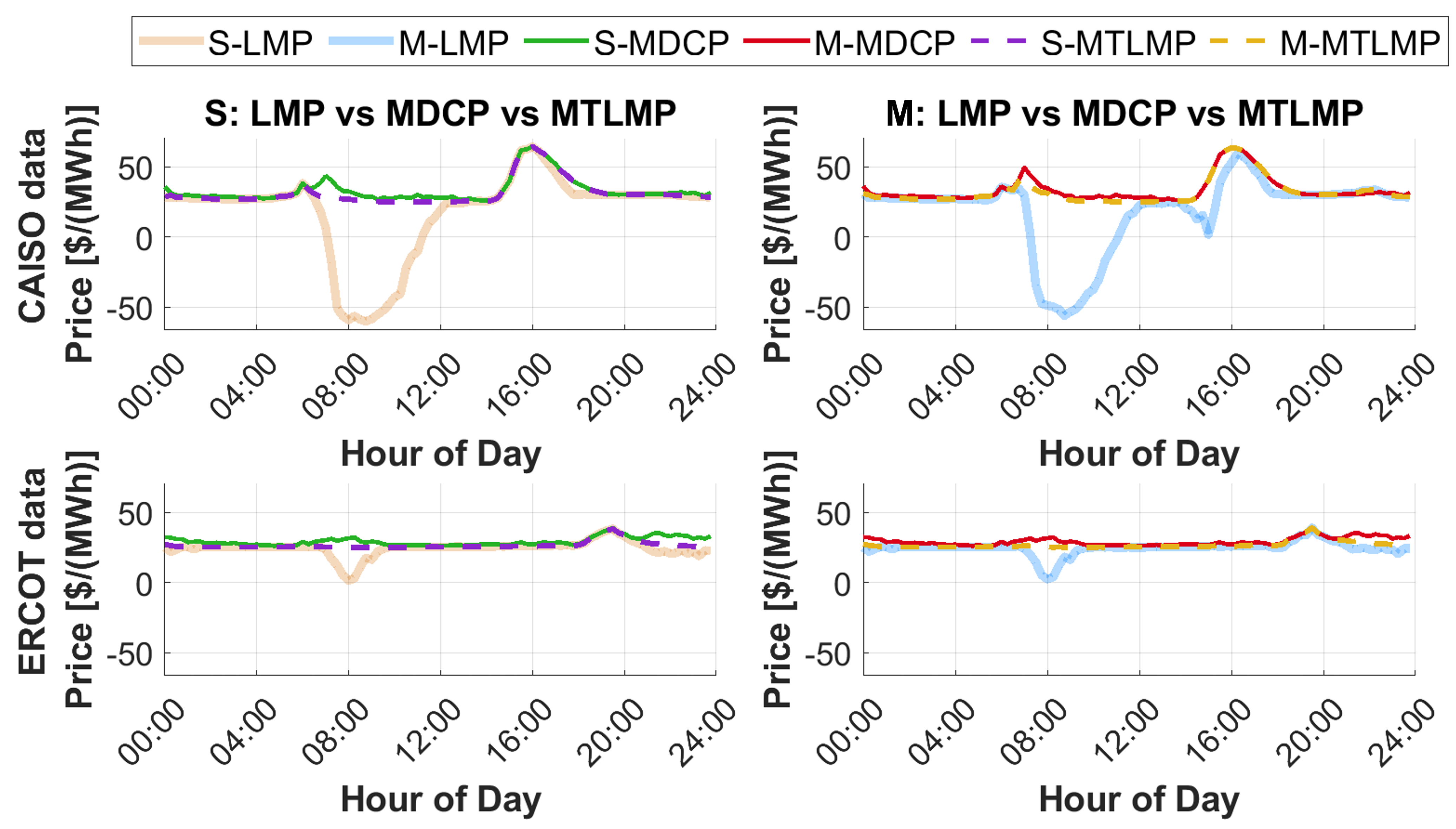}
    \vspace{-0.5cm}
    \caption{\scriptsize  100-day averaged hourly prices (LMP, MDCP, MTLMP) for ramping scenario S3, for CAISO and ERCOT  datasets. Prefixes S and M denote Single-interval and Multi-interval optimization formulations, respectively. }
    \label{fig:comparative_price_S3}
    %\vspace{-0.3cm}
\end{figure}

Fig.~\ref{fig:comparative_price_S3} compares hourly LMP, MDCP, and MTLMP averaged over 100 days' empirical datasets at CAISO and ERCOT. To distinguish between the dispatch models, pricing schemes under single-interval dispatch are denoted as S-LMP, S-MDCP, and S-MTLMP, while those under multi-interval dispatch are denoted as M-LMP, M-MDCP, and M-MTLMP.

The economic relationships and pricing behaviors observed in S3 were consistent across all ramping scenarios (S1--S10). As generators’ physical ramping capabilities increased beyond S3, ramping constraints became less binding, leading to more stable pricing signals and convergence across the different pricing mechanisms. Here we summarize key observations associated with S3 from Fig.~\ref{fig:comparative_price_S3}.

As a primary observation, the ERCOT dataset exhibited  significantly  lower price variability than CAISO. In the CAISO data, LMP showed substantially higher variations than both MDCP and MTLMP under single-interval and multi-interval dispatch. \tcb{Fig.~\ref{fig:price_variability} and Table~\ref{tab:overall_price_std} of the online appendix~\cite{chen2026ramping} report the price standard deviations.}  Specifically, the CAISO LMP signal is characterized by severe fluctuations; it dips sharply into negative values during the morning ramp-down, whereas MDCP and MTLMP effectively removed these deep price drops and appropriately increased prices during the steep afternoon ramp-up. Notably, negative LMP  in CAISO  reflected the effect of the curtailment of generation in the pricing signal and  led  to under-compensation and higher MWPs (see Sec.~\ref{sec:mwp}).

\tcb{To compare the pricing benchmarks, LMP is generally lower than MDCP and MTLMP. For example, during the morning ramp-down, procuring sufficient downward ramping capability requires dispatching more generators, which raises the highest marginal cost among dispatched units and thus raises MDCP. LMP does not rise, because the marginal energy can still be supplied by cheaper generators with available upward headroom. MDCP and MTLMP generally coincide under both single- and multi-interval dispatch, except during ramp-up periods, when MDCP is on average higher than MTLMP.} Interestingly, for the ERCOT dataset with lower net-demand volatility, the prices across all three methods (LMP, MDCP, and MTLMP) remained roughly the same.

When comparing each optimization approach, the relationship between multi-interval and single-interval prices depended on the timing of future ramping needs. Multi-interval prices were higher in the morning because a marginal increase in demand during the current binding interval exacerbated the difficulty of ramping down in subsequent intervals. Because multi-interval pricing explicitly considered both present and future operational costs, this anticipated future ramping-down scarcity was reflected as a higher price in the binding interval. In the sunset period, multi-interval prices can actually fall below single-interval prices. During periods of steep upward ramping, a marginal increase in generation in the current interval effectively pre-positioned the system, alleviating future ramping constraints and making future operations cheaper. Consequently, this anticipated future cost reduction translated into a lower overall marginal price for the multi-interval model compared to a myopic single-interval approach. 

It is important to note that while Fig.~\ref{fig:comparative_price_S3} reports the daily average prices—where MDCP and MTLMP generally appear to exceed or bound LMP—there is no strict mathematical ordering among the three pricing schemes. Depending on the specific binding constraints, generator bids, and system conditions within any individual dispatch interval, all possible ordinal relationships between LMP, MDCP, and MTLMP can materialize in practice.

\vspace*{-10 pt}%
\subsection{Generator Profits, MWP, and Demand Payments}\label{sec:mwp}

\begin{table*}[htbp]
    \centering
    \caption{Comparison of Generator Profit across Ramping Scenarios and Pricing Methods (Unit: \$)}
    \label{tab:profit_comparison}
    
    % --- Left Table: CAISO ---
    \begin{subtable}{0.48\textwidth}
        \centering
        \caption{CAISO Data}
        \label{tab:profit_caiso}
        \resizebox{\textwidth}{!}{%
        \begin{tabular}{l ccc ccc}
            \toprule
            & \multicolumn{3}{c}{\textbf{Single}} & \multicolumn{3}{c}{\textbf{Multi}} \\
            \cmidrule(lr){2-4} \cmidrule(lr){5-7}
            \textbf{\begin{tabular}{@{}l@{}}Ramp \\ Scenario\end{tabular}} & LMP & MTLMP & MDCP & LMP & MTLMP & MDCP \\
            \midrule
            S1 & 74587 & 85530 & 104498         & 82889 & 96389 & \textbf{107004} \\
            S3 & 61035 & 64865 & \textbf{67992} & 48325 & 57796 & 57446 \\
            S5 & 32273 & 32316 & \textbf{33629} & 32269 & 32372 & 33323 \\
            S8 & 30268 & 30268 & \textbf{30436} & 30268 & 30268 & 30436 \\
            \bottomrule
        \end{tabular}%
        }
    \end{subtable}%
    \hfill
    % --- Right Table: ERCOT ---
    \begin{subtable}{0.48\textwidth}
        \centering
        \caption{ERCOT Data}
        \label{tab:profit_ercot}
        \resizebox{\textwidth}{!}{%
        \begin{tabular}{l ccc ccc}
            \toprule
            & \multicolumn{3}{c}{\textbf{Single}} & \multicolumn{3}{c}{\textbf{Multi}} \\
            \cmidrule(lr){2-4} \cmidrule(lr){5-7}
            \textbf{\begin{tabular}{@{}l@{}}Ramp \\ Scenario\end{tabular}} & LMP & MTLMP & MDCP & LMP & MTLMP & MDCP \\
            \midrule
            S1 & 19468 & 20996 & \textbf{45800} & 20857 & 23222 & 44521 \\
            S3 & 15795 & 17353 & 17405          & 15469 & \textbf{17626} & 15903 \\
            S5 & 8426  & 8426  & \textbf{8471}  & 8426  & 8426  & 8471  \\
            S8 & 8343  & 8343  & \textbf{8343}  & 8343  & 8343  & 8343  \\
            \bottomrule
        \end{tabular}%
        }
    \end{subtable}
\end{table*}

\begin{table*}[htbp]
    \centering
    \caption{Comparison of MWP and Percentage Relative to Pure Energy Revenue across Ramping Scenarios. The percentages in parentheses represent the ratio of MWP to pure energy revenue without considering MWP. (Unit: \$)}
    \label{tab:mwp_comparison}
    
    % --- Left Table: CAISO ---
    \begin{subtable}{0.48\textwidth}
        \centering
        \caption{CAISO Data}
        \label{tab:mwp_caiso}
        \resizebox{\textwidth}{!}{%
        \begin{tabular}{l ccc ccc}
            \toprule
            & \multicolumn{3}{c}{\textbf{Single}} & \multicolumn{3}{c}{\textbf{Multi}} \\
            \cmidrule(lr){2-4} \cmidrule(lr){5-7}
            \textbf{\begin{tabular}{@{}l@{}}Ramp \\ Scenario\end{tabular}} & LMP & MTLMP & MDCP & LMP & MTLMP & MDCP \\
            \midrule
            S1 & 99561 (41.1\%) & 335 (0.1\%) & \textbf{0} (0.0\%) & 97773 (38.8\%) & 240 (0.1\%) & \textbf{0} (0.0\%) \\
            S3 & 5643 (1.7\%)   & 56 (0.0\%)  & \textbf{0} (0.0\%) & 6517 (2.1\%)   & 52 (0.0\%)  & \textbf{0} (0.0\%) \\
            S5 & 9127 (3.1\%)   & 34 (0.0\%)  & \textbf{0} (0.0\%) & 6527 (2.2\%)   & 27 (0.0\%)  & \textbf{0} (0.0\%) \\
            S8 & 2 (0.0\%)      & 2 (0.0\%)   & \textbf{0} (0.0\%) & 15 (0.0\%)     & 2 (0.0\%)   & \textbf{0} (0.0\%) \\
            \bottomrule
        \end{tabular}%
        }
    \end{subtable}%
    \hfill
    % --- Right Table: ERCOT ---
    \begin{subtable}{0.48\textwidth}
        \centering
        \caption{ERCOT Data}
        \label{tab:mwp_ercot}
        \resizebox{\textwidth}{!}{%
        \begin{tabular}{l ccc ccc}
            \toprule
            & \multicolumn{3}{c}{\textbf{Single}} & \multicolumn{3}{c}{\textbf{Multi}} \\
            \cmidrule(lr){2-4} \cmidrule(lr){5-7}
            \textbf{\begin{tabular}{@{}l@{}}Ramp \\ Scenario\end{tabular}} & LMP & MTLMP & MDCP & LMP & MTLMP & MDCP \\
            \midrule
            S1 & 14853 (6.5\%) & 389 (0.2\%) & \textbf{0} (0.0\%) & 14767 (6.4\%) & 349 (0.1\%) & \textbf{0} (0.0\%) \\
            S3 & 104 (0.0\%)   & 2 (0.0\%)   & \textbf{0} (0.0\%) & 183 (0.1\%)   & 2 (0.0\%)   & \textbf{0} (0.0\%) \\
            S5 & 1 (0.0\%)     & 1 (0.0\%)   & \textbf{0} (0.0\%) & 1 (0.0\%)     & 1 (0.0\%)   & \textbf{0} (0.0\%) \\
            S8 & 0 (0.0\%)     & 0 (0.0\%)   & \textbf{0} (0.0\%) & 0 (0.0\%)     & 0 (0.0\%)   & \textbf{0} (0.0\%) \\
            \bottomrule
        \end{tabular}%
        }
    \end{subtable}
\end{table*}

\begin{table*}[htbp]
    \centering
    \caption{Comparison of Demand Payment across Ramping Scenarios and Pricing Methods (Unit: \$)}
    \label{tab:demand_payment_comparison}
    
    % --- Left Table: CAISO ---
    \begin{subtable}{0.48\textwidth}
        \centering
        \caption{CAISO Data}
        \label{tab:dempay_caiso}
        \resizebox{\textwidth}{!}{%
        \begin{tabular}{l ccc ccc}
            \toprule
            & \multicolumn{3}{c}{\textbf{Single}} & \multicolumn{3}{c}{\textbf{Multi}} \\
            \cmidrule(lr){2-4} \cmidrule(lr){5-7}
            \textbf{\begin{tabular}{@{}l@{}}Ramp \\ Scenario\end{tabular}} & LMP & MTLMP & MDCP & LMP & MTLMP & MDCP \\
            \midrule
            S1 & \textbf{357183} & 368127          & 387095 & 364883          & 378384          & 388999 \\
            S3 & 337401          & 341231          & 344358 & \textbf{324231} & 333702          & 333352 \\
            S5 & \textbf{307208} & 307251          & 308564 & 307213          & 307316          & 308266 \\
            S8 & \textbf{305092} & \textbf{305092} & 305259 & \textbf{305092} & \textbf{305092} & 305259 \\
            \bottomrule
        \end{tabular}%
        }
    \end{subtable}%
    \hfill
    % --- Right Table: ERCOT ---
    \begin{subtable}{0.48\textwidth}
        \centering
        \caption{ERCOT Data}
        \label{tab:dempay_ercot}
        \resizebox{\textwidth}{!}{%
        \begin{tabular}{l ccc ccc}
            \toprule
            & \multicolumn{3}{c}{\textbf{Single}} & \multicolumn{3}{c}{\textbf{Multi}} \\
            \cmidrule(lr){2-4} \cmidrule(lr){5-7}
            \textbf{\begin{tabular}{@{}l@{}}Ramp \\ Scenario\end{tabular}} & LMP & MTLMP & MDCP & LMP & MTLMP & MDCP \\
            \midrule
            S1 & \textbf{247192} & 248720          & 273523          & 248545          & 250910          & 272210 \\
            S3 & 243031          & 244589          & 244641          & \textbf{242642} & 244799          & 243076 \\
            S5 & \textbf{235402} & \textbf{235402} & 235446          & \textbf{235402} & \textbf{235402} & 235446 \\
            S8 & \textbf{235312} & \textbf{235312} & \textbf{235312} & \textbf{235312} & \textbf{235312} & \textbf{235312} \\
            \bottomrule
        \end{tabular}%
        }
    \end{subtable}
    \vspace{-0.5cm}
\end{table*}

We summarize the results for average daily  generator profits (Table \ref{tab:profit_comparison}), MWP (Table \ref{tab:mwp_comparison}), and demand payments (Table \ref{tab:demand_payment_comparison}). 

%, across varying ramping capabilities for both the CAISO and ERCOT datasets. 

As the physical system became more constrained (moving from scenario S8 towards S1), we found that the magnitude of profits, MWP fractions, and payments noticeably increased, reflecting the higher operational difficulty and tighter constraints in both CAISO and ERCOT cases. On the other hand, as the system became less constrained (moving towards S8), the total profits and payments under the different pricing schemes converged with each other, mirroring the stabilization and convergence observed in the pricing signals themselves. 

Generator profit trends (Table~\ref{tab:profit_comparison}), including profits from energy, FRP, and MWP, highlight the impact of the different pricing mechanisms. Profits with LMP dispatch were consistently lower than those under MDCP and MTLMP. While MDCP generally resulted in the highest generator profits across most scenarios, we noted a specific exception in the ERCOT case under scenario S3, where M-MTLMP yielded the highest profit.

%This shows that given that there is no strict, absolute mathematical ordering between the clearing prices of MTLMP and MDCP, any relative ordering is possible depending on the specific binding constraints of the interval.

 \tcb{Differences in generator profit were closely tied to the MWPs reported in Table~\ref{tab:mwp_comparison}. MWPs were zero under MDCP in every scenario, consistent with Proposition~\ref{prop:minDemand}. \tcb{To explicitly highlight the severity of out-of-market uplifts, the percentages shown in parentheses \tcb{in Table~\ref{tab:mwp_comparison}} represent the MWP divided by the pure energy revenue (calculated without considering the MWP). Under MTLMP, they were 0.0\% of generators' total energy revenue in every scenario except S1, an extreme ramp-scarcity stress test, where they were only 0.1--0.2\%, consistent with Proposition~\ref{prop:MTLMPMW}. These nonzero MWPs arose only under pure ramp-down scheduling ($g^*_{it}=\underline{r}^*_{it}$) with a positive downward ramping price.}} In contrast, MWPs under LMP were significantly higher. We present interval-based MWP in the main text and daily MWP in the online appendix  \cite{chen2026ramping}. Both types of MWPs were comparable to or even larger than generator profits when ramping capabilities were highly constrained, like in Case S1 with frequent negative LMPs. Comparing the datasets, the severe variations in price in the CAISO data caused a massive portion of the total energy revenue to come from MWPs (reaching up to 41.1\% in S1). {\tcb{S1 is a stress test with extreme ramp scarcity, so its MWPs exceed those of the other scenarios and should not be read as typical of normal operations.}} This highlights the high frequency and severity of negative or excessively low LMPs in that case. In contrast, the relative MWP fraction of the energy revenue in the ERCOT case was significantly smaller.

Table~\ref{tab:demand_payment_comparison} illustrates the demand payments (including MWPs), showing that LMP consistently yields the lowest demand payments. In contrast, MDCP and MTLMP have higher demand payments because they internalize the out-of-market MWPs into the uniform in-market price. With limited ramping capacities in Case S1, MDCP added approximately 10\% more to the demand payment compared with MTLMP, while the change of MWP was small. This showed MDCP paid a high demand price to completely remove all MWPs in Propositions~\ref{prop:minDemand}. Comparatively, MTLMP in Propositions~\ref{prop:MTLMPMW} can mostly remove MWP while yielding a lower demand payment. Although this increased the upfront cost to consumers, incorporating these costs directly into the uniform market clearing price greatly improves market transparency. \tcb{It reflects the cost of ramping-constrained dispatch in the  price and eliminates the out-of-market uplifts that distort bidding incentives.}
%mention that only in S3 in ERCOT case, MTLMP becomes the pricing scheme with the highest gen profit, in the rest is MDCP. This is because there is not a strict order between the prices (all of the possible combinations are possible).

%\vspace*{-10 pt}%
\subsection{Multi-interval vs. single interval dispatch}\label{sec:model_comparison}

\begin{figure}[htbp]
    \centering
    % Top Image
     \vspace*{-10 pt}%
    \includegraphics[width=0.45\textwidth]{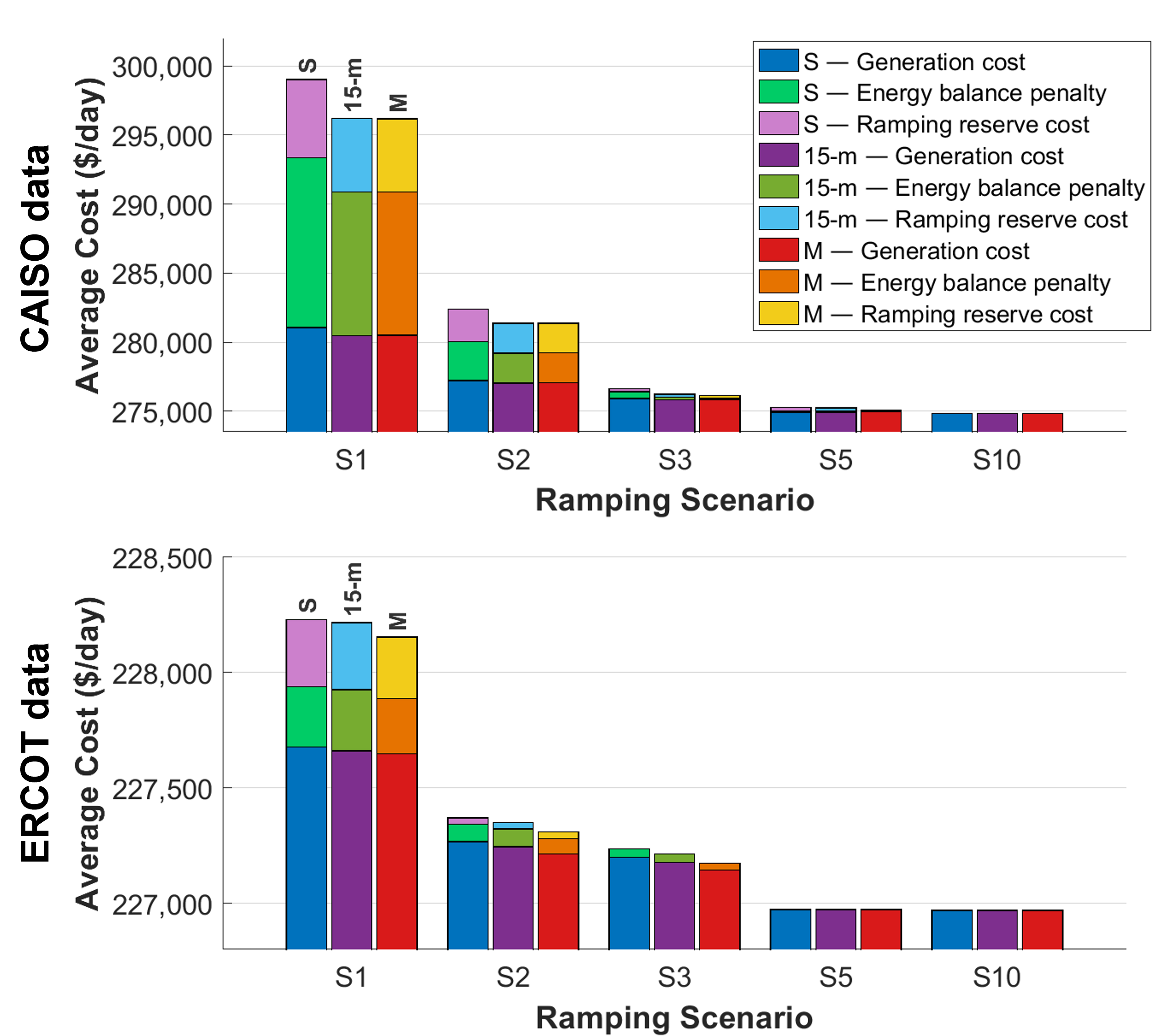}
    \caption{\scriptsize Daily average system operating cost decomposition for CAISO and ERCOT data across ramping scenarios S1-S10. The bars compare Single-interval (S), 15-minute uncertainty Multi-interval (15-m), and actual uncertainty Multi-interval (M) dispatch.}
    \label{fig:costs}
    \vspace{-0.3cm}
\end{figure}

%\tcr{-The average day with CAISO data is more challenging in terms of ramping requirements, making the average day to be more expensive that with the ERCOT case.
%-The fixed uncertainty treatment has also more expensive dispatch costs and is better than the single interval approach because we are still using the net demand forecast, energy procurement (the benefits of still being a multi-interval approach). Because of the mistaken form of treatment of uncertainty, it is more expensive than M approach.
%-In CAISO case, the system is still very challenged by the ramping events. I think that’s the main reason why the results with the fixed (F) treatment are comparable with the M treatment.
%Average ERCOT case total cost is significantly lower, compared to CAISO case.
%In terms of ordering, multi-interval dispatch with varying uncertainty remains as the cheapest option.
%Fixed uncertainty treatment in both cases is roughly the same to multi-interval with increasing uncertainty. 

%This subsection empirically compared multi-interval dispatch and single interval dispatch independently of the pricing method. Based on the binding-interval objective values of \eqref{eq:ED}, Fig.~\ref{fig:costs} shows the system operating cost including generation cost, ramping reserve costs, and penalties for energy imbalance (energy reserve and curtailment of renewables) are also computed from the objective of \eqref{eq:ED}. Lower system operating costs represent higher efficiencies.

This subsection provides an empirical comparison of operational efficiency (total cost) between single and multiple interval flexible ramp dispatch models, independently of the pricing method. Based on the binding-interval objective values of \eqref{eq:ED}, Fig.~\ref{fig:costs} shows the system operating cost including generation cost, ramping reserve costs (for ramp up and ramp down shortfalls), and penalties for energy imbalance (load shedding and renewable curtailment). Lower system operating costs represent higher efficiencies. Variations in total operating cost directly reflected differences in the quantities of energy imbalance, ramping shortfall, and reserve utilization procured in real time, because the energy imbalance penalty price and ramping reserve prices were exogenously fixed in simulation.

Current FRP implementations often rely on short-term uncertainty margins $\overline{U}_{t}$ and $\underline{U}_{t}$ (top left of Fig.~\ref{fig:RWramp}) that do not fully capture the increasing variance of net load over longer horizons. As noted by CAISO, the ``15-minute'' uncertainty used in current FRP designs is substantially smaller than the actual uncertainty that materializes over a multi-interval horizon \cite{CAISO:25FRP_DMM}. To evaluate the impact of uncertainty margin modeling and isolate the economic value of accurately capturing the growth of uncertainty over time, as opposed to merely incorporating a multi-interval horizon, we evaluated three frameworks: the myopic Single-interval (S) approach, a 15-minute (15-m) uncertainty multi-interval approach, and the fully dynamic actual uncertainty Multi-interval (M) approach. In M approach, flexible ramping requirements expand across later advisory intervals as forecast variance increases (details in Sec.~\ref{sec:forecastmain}). In contrast, the 15-m approach determines the uncertainty margin based solely on the immediate next interval and applies this same margin uniformly across all subsequent look-ahead intervals. Observations are summarized as follows.
%%%%

% We evaluated two variations of the multi-interval model: the fully dynamic uncertainty margin approach (denoted as M), where uncertainty margins naturally expand over the look-ahead horizon using the formulas detailed in Appendix~\ref{sec:demandforecast}; and the ``15-Minute'' uncertainty approach (denoted as 15-m), which statically uses the ramping requirement calculated for the binding interval across all future advisory intervals (see Section~\ref{sec:model_comparison} for further details). %Pricing schemes under the dynamic multi-interval dispatch are denoted as M-LMP, M-MDCP, and M-MTLMP. 

%%%
First, the majority of operating costs arose from baseline generation rather than penalty terms across all the scenarios. However, the efficiency gap between multi- and single-interval dispatch was primarily driven by penalties associated with energy imbalances and ramping shortfalls, particularly in tightly constrained scenarios (S1–S2). Moreover, evaluating the differences between the datasets, CAISO experienced more severe ramping events, resulting in significantly higher average operating costs. In contrast, ERCOT exhibited lower total costs due to the less physically demanding net-load profile and less challenging intra-day variability.

Second, when the system possessed sufficient physical flexibility to absorb net-load volatility (S5-S10), single- and multi-interval dispatch exhibited nearly identical performance. In these scenarios, the advanced preparation provided by the multi-interval look-ahead offered little to no additional value, and a single-interval dispatch was sufficient to maintain feasibility without triggering severe shortages.

Third, focusing on the differences between the forecasting frameworks, under highly constrained conditions, the performance of the 15-minute (15-m) uncertainty approach is notable. By incorporating a look-ahead horizon, it avoided the short-sightedness of single-interval dispatch and was consistently more cost-effective than the single-interval (S) approach. However, because it relied on a static representation of uncertainty, it failed to capture the growth in forecast variance across future intervals and thus remained more expensive than the fully actual uncertainty (M) approach. Overall, multi-interval dispatch with actual uncertainty (M) achieved the lowest cost. However, in the CAISO case, the system was so heavily challenged by steep physical ramping events that the primary economic benefit came simply from having any look-ahead horizon. As a result, the cost difference between (15-m) and (M) was relatively small.

Overall, when ramping requirements were properly specified, multi-interval dispatch consistently reduced total operating costs and improved system efficiency relative to single-interval dispatch. This finding is consistent with insights from stochastic control and with \cite{cavicchi18ramp}: maintaining ramp capability requires higher production costs in the current \textit{binding} interval, which is economically justified only when the expected future value of that ramp capability (e.g., avoiding a penalty) exceeds the immediate cost. %The effectiveness of multi-interval dispatch critically depends on the expected uncertainty—the more accurately the system operator anticipates when additional ramping capability will yield significant benefits, the more cost-effective the dispatch becomes.

%\vspace{-0.2cm}

\section{Conclusions}\label{sec:Conclusion}
High renewable penetration has intensified uncertainty and ramping requirements in the real-time electricity market, challenging existing dispatch and pricing mechanisms. This paper studies single- and multi-interval energy–ramping co-optimized real-time market with a focus on bid cost recovery (BCR), truthful bidding, and  operational efficiency. We propose two uniform pricing mechanisms—max dispatch cost pricing (MDCP) and max temporal locational marginal pricing (MTLMP). We theoretically show that MDCP eliminates BCR out-of-market uplifts and preserve truthful bidding incentives for price-taking generators.  We compare single- and multi-interval dispatch efficiency with ramping procurement under various forecast settings.

Empirically, we evaluate single- and multi-interval dispatch with LMP, MDCP, and MTLMP. Our results show that: (i) LMP is highly volatile and often negative in severe ramping scenarios, whereas MDCP and MTLMP eliminate negative prices and out of market BCR, yielding higher generator profits but higher demand payments; (ii) multi-interval dispatch internalizes future opportunity costs, leading to higher energy prices during constrained ramp-down periods to prevent future scarcity, while potentially lowering prices during ramp-up  by anticipating increased supply availability;  (iii) removing BCR uplifts improves pricing transparency but increases demand payments; and (iv) the relative efficiency of dispatch models depends critically on system flexibility and forecast accuracy: although dynamic multi-interval dispatch generally minimizes total operating costs in tightly constrained grids by pre-positioning resources, simpler industry practices—such as single-interval or fixed-margin multi-interval dispatch—are highly competitive alternatives that perform just as efficiently in systems with high ramping flexibility or less challenging net demand profiles. 

\tcb{Several directions remain open. First, the analysis uses a copper-plate model: MTLMP extends directly to transmission-constrained networks, whereas MDCP requires the optimal pricing formulation in the online appendix~\cite{chen2026ramping}; a closed-form extension and its truthful-bidding properties are under study~\cite{norambuenaguzman2026optimal}. Second, we do not model the different time granularities of RTUC and RTED in procuring energy and ramping capability. Third, our formulation couples generator minimum output limits with downward ramp capability; alternative formulations that decouple these constraints could mitigate potential under-compensation. Fourth, our incentive results concern short-run bidding: they assume price-taking generators and do not rule out gains for bidders that know the forecast-error distribution, and whether truthful bidding remains locally optimal in either case is open. Long-run incentives to invest in ramping capability require a model of investment decisions and are also left for future work. Fifth, the framework could be extended beyond electricity-only markets: to multi-energy markets and to dispatch that accounts for the fast dynamics of inverter-based resources~\cite{She2024VIS}. Finally, although the IEEE RTS-24 results in the online appendix qualitatively agree with those of the three-generator system, the empirical ranking of LMP, MDCP, and MTLMP may vary across systems, so system-specific conclusions require simulations of the system in question.}

{
\bibliographystyle{IEEEtran}
\bibliography{BIB}

@article{She2024VIS,
  author  = {She, Buxin and Li, Fangxing and Cui, Hantao and Wang, Jinning and Zhang, Qiwei and Bo, Rui},
  title   = {Virtual Inertia Scheduling ({VIS}) for Real-Time Economic Dispatch of {IBR}-Penetrated Power Systems},
  journal = {IEEE Transactions on Sustainable Energy},
  volume  = {15},
  number  = {2},
  pages   = {938--951},
  year    = {2024},
  month   = apr
}

@article{Brown2010,
  author  = {Brown, David B. and Smith, James E. and Sun, Peng},
  title   = {Information Relaxations and Duality in Stochastic Dynamic Programs},
  journal = {Operations Research},
  volume  = {58},
  number  = {4},
  pages   = {785--801},
  year    = {2010},
  doi     = {10.1287/opre.1090.0796}
}

@misc{norambuenaguzman2026optimal,
  title        = {Optimal Uniform Pricing for Multi-Interval Dispatch without Make-Whole Uplifts},
  author       = {Norambuena-Guzman, Valentina and Chen, Cong and Tong, Lang and Mount, Timothy D.},
  year         = {2026},
  month        = sep,
  eprint       = {2609.00541},
  archivePrefix= {arXiv},
  primaryClass = {eess.SY},
  url          = {https://arxiv.org/abs/2609.00541}
}

@article{Conejo:02TPS,
  author  = {A. J. Conejo and F. J. Nogales and J. M. Arroyo},
  title   = {Price-taker bidding strategy under price uncertainty},
  journal = {IEEE Trans. Power Syst.}, volume = {17}, number = {4}, pages = {1081--1088}, year = {2002}}

@article{KAHN200170,
title = {Uniform Pricing or Pay-as-Bid Pricing: A Dilemma for California and Beyond},
journal = {The Electricity Journal},
volume = {14},
number = {6},
pages = {70-79},
year = {2001},
issn = {1040-6190},
doi = {https://doi.org/10.1016/S1040-6190(01)00216-0},
url = {https://www.sciencedirect.com/science/article/pii/S1040619001002160},
author = {Alfred E. Kahn and Peter C. Cramton and Robert H. Porter and Richard D. Tabors}
}

@article{Hogan14EJ,
  author  = {W. W. Hogan},
  title   = {Electricity market design and efficient pricing: Applications for {New England} and beyond},
  journal = {The Electricity Journal}, volume = {27}, number = {7}, pages = {23--49}, year = {2014}}

@book{Schweppe:88,
  author    = {F. C. Schweppe and M. C. Caramanis and R. D. Tabors and R. E. Bohn},
  title     = {Spot Pricing of Electricity},
  publisher = {Kluwer Academic Publishers}, address = {Boston, MA}, year = {1988}}

@misc{FERCJPM,
  author       = {{Federal Energy Regulatory Commission}},
  title        = {Order Approving Stipulation and Consent Agreement, {JP Morgan Ventures Energy Corporation}},
  howpublished = {144 FERC \P~61,068, Docket No. IN11-8-000},
  note         = {Jul. 30, 2013},
  year         = {2013},
  url          = {https://www.ferc.gov/sites/default/files/EventCalendar/Files/20130730080931-IN11-8-000.pdf}
}

@article{wilson2002architecture,
  title   = {Architecture of Power Markets},
  author  = {Wilson, Robert},
  journal = {Econometrica},
  volume  = {70},
  number  = {4},
  pages   = {1299--1340},
  year    = {2002},
  month   = jul,
  doi     = {10.1111/1468-0262.00334}
}

@techreport{ordoudis2016updated,
  author      = {Christos Ordoudis and Pierre Pinson and Juan M. Morales and Marco Zugno},
  title       = {An Updated Version of the {IEEE RTS} 24-Bus System for Electricity Market and Power System Operation Studies},
  institution = {Technical University of Denmark},
  address     = {Kgs. Lyngby, Denmark},
  year        = {2016}
}

@article{chen2026ramping,
  title={Appendix: Ramping Procurement and Bid-Cost Recovery in Real-Time Market},
  author={Chen, Cong and Norambuena, Valentina and Tong, Lang},
  journal={arXiv preprint arXiv:2606.19599},
  year={2026}
}

@inproceedings{Werner23CDCpricing,
  title={Pricing uncertainty in stochastic multi-stage electricity markets},
  author={Werner, Lucien and Christianson, Nicolas and Zocca, Alessandro and Wierman, Adam and Low, Steven},
  booktitle={2023 62nd IEEE Conference on Decision and Control (CDC)},
  pages={1580--1587},
  year={2023},
  organization={IEEE}
}

@inproceedings{Zhang26Ramp,
  author    = {Zhang, Q. and Xie, L. and Zhao, L. and Wang, C.},
  title     = {Comparative Assessment of Look-Ahead Economic Dispatch and Ramp Products for Grid Flexibility},
  booktitle = {Proceedings of the Power Systems Computation Conference (PSCC)},
url = {https://arxiv.org/pdf/2601.22120},
  year      = {2026}
}

@misc{CAISO:24BCR,
  title =        {Storage bid cost recovery and default energy bids enhancements},
  howpublished = "[ONLINE], available  (2025/02/11) at \url{https://stakeholdercenter.caiso.com/StakeholderInitiatives/storage-bid-cost-recovery-and-default-energy-bids-enhancements}",
  year =         {2024},
  month =        {November}
}

@article{Cho:23OR,
  title={Pricing under uncertainty in multi-interval real-time markets},
  author={Cho, Jehum and Papavasiliou, Anthony},
  journal={Operations research},
  volume={71},
  number={6},
  pages={1928--1942},
  year={2023},
  publisher={INFORMS}
}

@misc{Hogan:20,
  title={Electricity market design: Multi-interval pricing models},
  author={Hogan, William W},
  howpublished = "[ONLINE], available  (2025/02/11) at \url{https://scholar.harvard.edu/files/whogan/files/hogan_hepg_multi_period_062220.pdf}",
  year={2020},
  month =        {June}
  }

@article{Biggar:22EJ,
  title={Do we need to implement multi-interval real-time markets?},
  author={Biggar, Darryl R and Hesamzadeh, Mohammad Reza},
  journal={The Energy Journal},
  volume={43},
  number={2},
  pages={111--131},
  year={2022},
  publisher={SAGE Publications Sage CA: Los Angeles, CA}
}

@article{Mays:24EE,
  title={Sequential pricing of electricity},
  author={Mays, Jacob},
  journal={Energy Economics},
  volume={137},
  pages={107790},
  year={2024},
  publisher={Elsevier}
}

@article{Zhao&Zheng&Litvinov:19TPS,
  title={A multi-period market design for markets with intertemporal constraints},
  author={Zhao, Jinye and Zheng, Tongxin and Litvinov, Eugene},
  journal={IEEE Transactions on Power Systems},
  volume={35},
  number={4},
  pages={3015--3025},
  year={2019},
  publisher={IEEE}
}

@article{Guo&Chen&Tong:21TPS,
  author={Guo, Ye and Chen, Cong and Tong, Lang},
  journal={IEEE Transactions on Power Systems}, 
  title={Pricing Multi-Interval Dispatch Under Uncertainty Part {I}: Dispatch-Following Incentives}, 
  year={2021},
  volume={36},
  number={5},
  pages={3865-3877},
  doi={10.1109/TPWRS.2021.3055730}}

@article{Chen&Guo&Tong:20TPS,
  title={Pricing multi-interval dispatch under uncertainty part {II}: Generalization and performance},
  author={Chen, Cong and Guo, Ye and Tong, Lang},
  journal={IEEE Transactions on Power Systems},
  volume={36},
  number={5},
  pages={3878--3886},
  year={2020},
  publisher={IEEE}
}

@ARTICLE{YurdakulEla25FRPDA,
  author={Yurdakul, Ogün and Ela, Erik and Billimoria, Farhad},
  journal={IEEE Transactions on Energy Markets, Policy and Regulation}, 
  title={Flexible Ramping Product Procurement in Day-Ahead Markets}, 
  year={2025},
  volume={3},
  number={1},
  pages={13-31},
  doi={10.1109/TEMPR.2024.3453907}}

@ARTICLE{GhaljeheiKhorsand22ADAFRP,
  author={Ghaljehei, Mohammad and Khorsand, Mojdeh},
  journal={IEEE Transactions on Power Systems}, 
  title={Day-Ahead Operational Scheduling With Enhanced Flexible Ramping Product: Design and Analysis}, 
  year={2022},
  volume={37},
  number={3},
  pages={1842-1856},
  doi={10.1109/TPWRS.2021.3110712}}

@article{Topkis78,
 ISSN = {0030364X, 15265463},
 URL = {http://www.jstor.org/stable/169636},
 author = {Donald M. Topkis},
 journal = {Operations Research},
 number = {2},
 pages = {305--321},
 publisher = {INFORMS},
 title = {Minimizing a Submodular Function on a Lattice},
 urldate = {2026-06-08},
 volume = {26},
 year = {1978}
}

@article{Tong&Wang:26NOW,
  title         = {{AI} Foundation Model for Time Series with Innovations Representation},
  author        = {Tong, Lang and Wang, Xinyi},
  journal       = {arXiv preprint arXiv:2510.01560},
  year          = {2025},
  eprint        = {2510.01560},
  archivePrefix = {arXiv},
  primaryClass  = {stat.ML},
  doi           = {10.48550/arXiv.2510.01560},
  url           = {https://arxiv.org/abs/2510.01560}
}

@INPROCEEDINGS{ZhangKory19rampuplift,
  author={Zhang, Shaobo and Hedman, Kory W.},
  booktitle={2019 North American Power Symposium (NAPS)}, 
  title={Conditions for Ramp Rates Causing Uplift}, 
  year={2019},
  volume={},
  number={},
  pages={1-6},
  doi={10.1109/NAPS46351.2019.9000202}}

@INPROCEEDINGS{ChenTong25PESGM,
  author={Chen, Cong and Tong, Lang},
  booktitle={2025 IEEE Power \& Energy Society General Meeting (PESGM)}, 
  title={Incentivizing Ramping with Uniform Pricing}, 
  year={2025},
  volume={},
  number={},
  pages={1-5},
  doi={10.1109/PESGM52009.2025.11225414}}

@misc{CAISO:20FRPRefinements,
  title        = {Flexible Ramping Product Refinements: Final Proposal},
  howpublished = "[ONLINE], available (2025/11/16) at \url{https://stakeholdercenter.caiso.com/initiativedocuments/finalproposal-flexiblerampingproductrefinements.pdf}",
  year         = {2020},
  month        = {August}
}

@misc{CAISO:23FRP,
  title        = {California {ISO} to {Chile}: Flexible Ramping Product},
  howpublished = "[ONLINE], available (2025/11/16) at \url{https://www.caiso.com/documents/presentation-california-iso-to-chile-flexible-ramping-product-jun-15-2023.pdf}",
  year         = {2023},
  month        = {June}
}

@misc{MISO:Schedule29,
  title        = {Dynamic Ramp Capability
Product Requirement (Uncertainty Component)},
  howpublished = "[ONLINE], available (2025/11/16)  at \url{https://cdn.misoenergy.org/%2020250828%20RSC%20Item%2009%20Dynamic%20Ramp%20Capability%20Product-Uncertainty%20Component%20(RSC%202024-2)715286.pdf}",
  year         = {2025}
}

@misc{CAISO:22BCR,
  title        = {Tariff Amendment to Prevent Unwarranted Bid Cost Recovery Payments to Storage Resources},
  howpublished = "[ONLINE], available (2025/11/16) at \url{https://www.caiso.com/Documents/Sep19-2022-TariffAmendment-EnergyStorageBidCostRecovery-ER22-2881.pdf}",
  year         = {2022},
  month        = {September}
}

@misc{CAISO:25BCRMIO,
  title =        {Storage Design and Modeling},
  howpublished = "[ONLINE], available  (2025/10/20) at \url{https://stakeholdercenter.caiso.com/InitiativeDocuments/Presentation-Storage-Design-and-Modeling-Sep-29-2025.pdf}",
  year =         {2025},
  month =        {September}
}

@misc{CAISO:11FRP,
  title =        {Flexible Ramping Products Straw Proposal},
  howpublished = "[ONLINE], available  (2025/12/07) at \url{https://www.caiso.com/documents/flexiblerampingproductstrawproposal.pdf}",
  year =         {2011},
  month =        {November}
}

@misc{FERC15upliftE-2_14,
  title =        {Price Formation in Energy and Ancillary Services Markets Operated by Regional Transmission Organizations and Independent System Operators},
  howpublished = "[ONLINE], available  (2025/12/16) at \url{https://www.ferc.gov/sites/default/files/2020-06/E-2_14.pdf}",
  year =         {2015},
  month =        {November}
}

@misc{Clyde24CAISOramp,
  title =        {Flexible Capacity Requirement Methodology for 2025 through 2027},
  howpublished = "[ONLINE], available at \url{https://www.google.com/url?sa=t&source=web&rct=j&opi=89978449&url=https://stakeholdercenter.caiso.com/InitiativeDocuments/Presentation-2025-Flexible-Capacity-Needs-Assessment-Feb122024.pdf&ved=2ahUKEwjXo8WXnsORAxXOGVkFHd5sCtQQFnoECBwQAQ&usg=AOvVaw0iA5Wslk7w2aO9wZtQV4Zf}",
  year =         {2024},
  month =        {February}
}

@article{wuhug15TPSfrp,
  title={Risk-limiting economic dispatch for electricity markets with flexible ramping products},
  author={Wu, Chenye and Hug, Gabriela and Kar, Soummya},
  journal={IEEE Transactions on Power Systems},
  volume={31},
  number={3},
  pages={1990--2003},
  year={2015},
  publisher={IEEE}
}

@ARTICLE{WuPapalexopoulos04TPSreservepricing,
  author={Tong Wu and Rothleder, M. and Alaywan, Z. and Papalexopoulos, A.D.},
  journal={IEEE Transactions on Power Systems}, 
  title={Pricing energy and ancillary services in integrated market systems by an optimal power flow}, 
  year={2004},
  volume={19},
  number={1},
  pages={339-347},
  doi={10.1109/TPWRS.2003.820701}}

@article{wanghobbs15TPSfrp,
  author={Wang, Beibei and Hobbs, Benjamin F.},
  journal={IEEE Transactions on Power Systems}, 
  title={Real-Time Markets for Flexiramp: A Stochastic Unit Commitment-Based Analysis}, 
  year={2016},
  volume={31},
  number={2},
  pages={846-860},
  doi={10.1109/TPWRS.2015.2411268}}

@article{ela15TPSrtFRP,
  title={Scheduling and pricing for expected ramp capability in real-time power markets},
  author={Ela, Erik and O'Malley, Mark},
  journal={IEEE Transactions on Power Systems},
  volume={31},
  number={3},
  pages={1681--1691},
  year={2015},
  publisher={IEEE}
}

@INPROCEEDINGS{cavicchi18ramp,
  title={Ramp capability dispatch and uncertain intermittent resource output},
  author={Cavicchi, Joseph and Harvey, Scott},
  booktitle={Rutgers Center for Research in Regulated Industries Advanced Workshop in Regulation and Competition 31st Annual Western Conference, Hyatt Regency, Monterey, California},
  year={2018}
}

@misc{CAISO:25FRP_DMM,
  title =        {Recommendation to increase the {FRP} uncertainty horizon},
  author =       {Kyle Westendorf},
  howpublished = "[ONLINE], available (2026/03/20) at \url{https://stakeholdercenter.caiso.com/InitiativeDocuments/Presentation-Department-of-Market-Monitoring-Flexibility-Ramping-Product-Jan-22-2025.pdf}",
  year =         {2025},
  month =        {January}
}

@misc{CAISOData,
  title =        {Today's Outlook},
  author =       {{California Independent System Operator}},
  howpublished = "[ONLINE], available (2026/10/27) at \url{https://www.caiso.com/todays-outlook}"
}

@misc{ERCOTLoadData,
  title =        {2-Day Real Time Gen and Load Data Reports},
  author =       {{Electric Reliability Council of Texas}},
  note =         {Product NP3-910-ER, Report Type 13056},
  howpublished = "[ONLINE], available (2026/10/27) at \url{https://www.ercot.com/mp/data-products/data-product-details?id=NP3-910-ER}"
}

@misc{ERCOTWindData,
  title =        {Wind Power Production---Actual 5-Minute Averaged Values},
  author =       {{Electric Reliability Council of Texas}},
  note =         {Product NP4-733-CD, Report Type 13071},
  howpublished = "[ONLINE], available (2026/10/27) at \url{https://www.ercot.com/mp/data-products/data-product-details?id=NP4-733-CD}"
}

@misc{ERCOTSolarData,
  title =        {Solar Power Production---Actual 5-Minute Averaged Values by Geographical Region},
  author =       {{Electric Reliability Council of Texas}},
  note =         {Product NP4-746-CD, Report Type 21810},
  howpublished = "[ONLINE], available (2026/10/27) at \url{https://www.ercot.com/mp/data-products/data-product-details?id=NP4-746-CD}"
}
}

% \newpage
\section{Appendix}\label{sec:Appendix}

 \subsection{Proof of Proposition~1}

We prove by showing MDCP is the optimizer of the following problem: 
 \begin{equation} \label{eq:mDPdemandallT}
\begin{array}{lrl}
&\underset{\pmb{\pi}=(\pi_{t})}{\rm minimize} &   \sum\limits_{t\in [T]} \pi_{t} d_{t} \\& {\rm subject~to} & \forall i \in [N], \forall t \in [T],\\& & {\cal M}_{it}(\pi_t, g^{\R}_{it}) = 0,\\
&&\pi_t \geq p_t \mathbbm{1}_{[s^{\R}_{t}>0]}.
\end{array}
\end{equation}
where we assume ex-post demand  $(d_t) \in \mathbb{R}^+, \forall t \in [T]$.

For a generator $i$ at time $t$, to satisfy ${\cal M}_{it}(\pi_t, g^{\R}_{it}) = 0$ defined by (7), we need $\pi_{t} \geq c_{it} $ if $g^{\R}_{it}>0$. So ${\cal M}_{it}(\pi_t, g^{\R}_{it}) = 0$, $\forall i \in [N], \forall t \in [T]$ if and only if $\pi_{t} \geq \underset{i  \in [N]}{\rm max}~~c_{it}\mathbbm{1}_{[g^{\R}_{it}>0]}.$ Therefore, constraints of \eqref{eq:mDPdemandallT} can be rewritten into
\beq\label{eq:feasible}
\pi_{t} \geq {\rm max}\{ \underset{i  \in [N]}{\rm max}~~c_{it}\mathbbm{1}_{[g^{\R}_{it}>0]}, \quad  p_t \mathbbm{1}_{[s^{\R}_{t}>0]}\}
\eeq
From the MDCP definition in (8), we know  \eqref{eq:feasible} is satisfied.  

Under the assumption that  $d_t \in \mathbb{R}^+$, we know the lower bound in \eqref{eq:feasible} is the optimizer of \eqref{eq:mDPdemandallT}, which is exactly MDCP defined in  (8). \hfill$\square$
 
\vspace{-0.3cm}
\subsection{Proof of Proposition~2}

Here we compare MTLMP and the marginal cost of a scheduled generator $i$ at time $t$. Denote $\overline{\rho}_{it}$ and $\underline{\rho}_{it}$ as dual variables associated with two constraints in (1g), respectively. Denote $\bar{\gamma}_{it}^*, \underline{\gamma}^*_{i(t+1)}$ as dual variables associated with ramp up and ramp down constraints in (1e-1f).  Use superscript * for the optimal dual solution. From the stationary in KKT conditions of (1), we have
\beq
\begin{array}{lrl}
&& c_{it}-\lambda_t^*-\underline{\gamma}_{it}^*+\bar{\gamma}_{it}^*+\underline{\gamma}^*_{i(t+1)}-\bar{\gamma}^*_{i(t+1)}-\underline{\rho}^*_{it}+\bar{\rho}^*_{it}=0\\
 &&~~~~\overset{(a)}{\Rightarrow}   c_{it}-\pi^{\text{\tiny TLMP}}_{it} -\underline{\rho}^*_{it}+\bar{\rho}^*_{it}=0 \\
 &&~~~~\overset{(b)}{\Rightarrow}   \pi^{\TLMP}_{it} - c_{it} = \bar{\rho}^*_{it} \geq 0 \\
 &&~~~~\overset{(c)}{\Rightarrow}   \pi^{\mTLMP}_{t} \geq  \pi^{\text{\tiny TLMP}}_{it} \geq  c_{it} . \nn
 \end{array}
 \eeq
Here, (a) comes from replacing in the definition of  TLMP (Sec.\ref{eq:MLMPderive}) and (b) comes from the assumption that $g^*_{it} \neq \underline{r}^{*}_{it}$ and the complementary slackness condition $\underline{\rho}_{it}^*(g^*_{it} - \underline{r}^{*}_{it})=0$. When the generator is scheduled, i.e. $g^*_{it}>0$, this complimentary slackness condition gives that $\underline{\rho}_{it}^*=0$. The last step (c) comes from the definition of MTLMP in (10) guaranteed to be no less than TLMP. 

Now that we show that MTLMP is always no less than the marginal cost for a scheduled generator, we have ${\cal M}_{it}(\pi_t, g^{\R}_{it}) = 0,   \forall i \in [N], \forall t \in [T]$ under  MTLMP. \hfill$\square$
   
\subsection{Proof of Theorem~1}
 
We first provide explicit formulation for profit maximization from the perspective of a price-taking generator and then prove Theorem~1 by two steps.

Theorem~1  establishes  the truthful-bidding results for price-taking generators under MDCP, demonstrating that truthful bidding is a local Nash-equilibrium strategy. By a price-taking bidder, we mean a bidder who assumes that its own bidding action cannot influence the market shadow price (LMP). In particular, the shadow prices $\lambda_t^*$  of the power balance constraints (1b) are exogenously determined. The bid cost trajectory is denoted by $\mathbf{c}_i := (c_{it})_{t \in \mathcal{T}_{t'}}$.  When submitting its bid cost to the market-clearing process, generator $i$ treats the shadow price $\boldsymbol{\lambda}^*:=(\lambda_t^*)$  as given and chooses the optimal bid cost $\cbf_i^*$ to maximize its {\em estimated profit}, \ie
\beq\label{eq:profitQ}
%\begin{array}{lrl}
  \max_\cbf~~ Q_i(\cbf_i):= \sum_{t\in {\cal T}_{t'}}  (\hat{ \pi}^{\mDP}_t(\cbf_i)-\cbf^\dagger_i)\hat{g}_{it}(\mathbf{c}, \boldsymbol{\lambda}^*).
%\end{array}
\eeq
Here, $c_{it}^\dagger$ denote the true marginal cost. Under MDCP, generators receive zero out-of-market MWP, and the in-market profit is the only objective in \eqref{eq:profitQ} relevant for bidding decisions. %Theorem~\ref{thm:bidRW} establishes that, under MDCP, a price-taking generator locally maximizing the forward-looking profit   has no incentive to bid above its true marginal cost within a rolling window $\mathcal{T}_{t'}$. 
Note that while, by assumption,  the price taker cannot influence the shadow price, by definition of MDCP, it can influence MDCP indirectly. From \eqref{eq:profitQ}, the effect of bid-cost has two effects on profit: a higher bid $\cbf_i$ could  increase the estimated MDCP $\hat{ \pibf}^{\mDP}_t(\cbf)$. At the same time, it could also reduce the dispatched generation quantity $\hat{g}_{it}(\mathbf{c}, \boldsymbol{\lambda}^*)$, and possibly not be dispatched in interval $t'$.  $\hat{g}_{it}(\mathbf{c}, \boldsymbol{\lambda}^*)$ denotes the estimated dispatched quantity when price-taking generator is given shadow price $\boldsymbol{\lambda}^*$.  $\hat{g}_{it}(\mathbf{c}_i,\boldsymbol{\lambda}^*)$ is computed by \eqref{eq:iniviG} rather than $g_{it}^{\mathrm{\R}}(\mathbf{c}_i)$ obtained from the centralized problem (1)-(2), since individual generators do not possess complete system-level information required to reconstruct the centralized dispatch. Under dual decomposition, \eqref{eq:iniviG} yields the same optimal dispatch as (1) when FRPs  are fixed at the optimal. 

\subsubsection{Estimated MDCP} To analyze how MDCP changes when bid-in cost $\cbf_i$ changes, we establish the following relationship based on the definition of estimated MDCP from the perspective of price-taking generators
\beq\label{eq:MDCPchange}
\hat{ \pi}^{\mDP}_{t}(\cbf_i) =\begin{cases}c_{it}, &\quad \{\hat{g}_{it}>0 \}\cap \{c_{it} > \pi^{\mDP}_{-it} \},\\ \pi^{\mDP}_{-it} , &\quad \text{otherwise}, \end{cases}
\eeq
where $\pi^{\mDP}_{-it}$ is the MDCP at time $t$ when  generator $i$ is excluded from the optimization of economic dispatch (1), and $\hat{g}_{it}$ is computed by  \eqref{eq:iniviG}. Equation \eqref{eq:MDCPchange} follows MDCP definition in (8)  and depends only on information available to generator $i$. Under the price-taking assumption, the  shadow price of power balance constraint $\boldsymbol{\lambda}^*$ and the dispatch decisions of other generators remain unchanged when generator $i$ unilaterally deviates in its bid $\mathbf{c}_i$. A deviation of generator $i$ affects MDCP only in the event that it  get dispatched with a higher bid-in cost, \ie $\{\hat{g}_{it}>0 \}\cap \{c_{it} > \pi^{\mDP}_{-it} \}$.   Otherwise,  MDCP remain unchanged, implying $ \hat{ \pi}_{t}^{\mDP} = \pi^{\mDP}_{-it}$.

\subsubsection{Estimated dispatch} 
   By a price-taking bidder, we mean a bidder who assumes that its own bidding action cannot influence the market-clearing price. In particular, the shadow price of the power balance constraint $\lambda_t^*$ are exogenously determined by (1b). When submitting a bid as a price–quantity pair, the generator treats the market price as given and determines the quantity it is willing to supply at that price. Accordingly, the estimated profit-maximizing dispatch  of generator $i$ over the window $\mathcal{T}_{t'}$ is  given by
\begin{equation} \label{eq:iniviG}
\begin{array}{lrl}
&\hat{ \gbf}_i(\cbf_i,\boldsymbol{\lambda}^*):=\underset{\gbf_{i}}{\rm arg~max} & \sum_{t\in {\cal T}_{t'}}(\lambda_t^*  - c_{it}) g_{it},   \\
&{\rm subject~to} & \forall t \in {\cal T}_{t'}\\
&&-r^{\D}_{i}\le g_{it'}-g_{i(t'-1)} \le r^{\U}_{i}, \\
&& -\underline{r}^*_{it}\le g_{i(t+1)}-g_{it} \le \overline{r}^*_{it}, \\
&&\underline{r}^*_{it}\le g_{it}\le \overline{g}_i -\overline{r}^*_{it}.
\end{array}
\end{equation}
Here, we focus on price-taking generators under the optimal FRP procurement $\{\mathbf{\overline{r}}^*_t, \mathbf{\underline{r}}^*_t\}_{t\in \mathcal{T}_{t'}}$ within the rolling window $\mathcal{T}_{t'}$.  The bid cost trajectory is denoted by $\mathbf{c}_i := (c_{it})_{t \in \mathcal{T}_{t'}}$. The constraints include the initial ramping condition, intertemporal ramping limits, and capacity constraints adjusted for procured FRP.  While in practice the bidder may have side information about other bidders' actions or the expected market price, the theoretical analysis here excludes such information and considers the optimal strategy to exogenously given prices.  We focus on the non-degenerate case in which the optimal solution to \eqref{eq:iniviG} is unique.

\subsubsection{Proof of Theorem~1} 

We prove Theorem~1 by two steps.   

\underline{Step 1.} We first show that the profit of generator $i$ at time interval $t$ in \eqref{eq:profitQ} has the following equivalence 
\begin{equation} \label{eq:profitEQ}
\begin{array}{lrl}
Q_{i}(\cbf_i)&:=& \sum_{t\in {\cal T}_{t'}}   (\hat{ \pi}^{\mDP}_{t}(\cbf_i)  - c_{it}^\dagger) \hat{g}_{it}(\cbf_i,\boldsymbol{\lambda}^*)\\
&=& \sum_{t\in {\cal T}_{t'}}   (\overline{c}_{it}  - c_{it}^\dagger) \hat{g}_{it}(\cbf_i,\boldsymbol{\lambda}^*). \\
\end{array}
\end{equation}
Notation $\hat{ \pi}^{\mDP}_{t}(\cbf_i)$  makes it explicit for the dependency of estimated MDCP $\hat{ \pi}^{\mDP}_{t}$ on generator i’s bid $\cbf_i$. By definition, we have $\overline{\cbf}_i:=\max\{\cbf^\dagger_{i}, \boldsymbol{\pi}^{\mDP}_{-i}\}$,
 \begin{itemize}
     \item \underline{If $c^\dagger_{it}<\pi^{\mDP}_{-it}$,} then   $ \cbf_i \in {\cal C}_i:=\{\cbf: \cbf^\dagger_{i} \le \cbf \le \overline{\cbf}_i\}$ implies  
     $$
     \hat{ \pi}^{\mDP}_{t}(\cbf_i) \overset{(a)}{=}\pi^{\mDP}_{-it}\overset{(b)}{=}\overline{c}_{it}.$$
     Here, (a) comes from definition of MDCP with more explanations in \eqref{eq:MDCPchange}  and (b) comes from the definition of $\overline{\cbf}_i$.
     Thus \eqref{eq:profitEQ} holds.
      \item \underline{If $c^\dagger_{it} \geq \pi^{\mDP}_{-it}$,} then by definition $\overline{c}_{it}=c^\dagger_{it}$ and $ \cbf_i \in {\cal C}_i$ implies  $c_{it}=c^\dagger_{it}$. If $\hat{g}_{it}=0$ solved by \eqref{eq:iniviG}, then \eqref{eq:profitEQ} holds and equals 0. If $\hat{g}_{it}>0$, then from the MDCP definition  in \eqref{eq:MDCPchange}, $\hat{ \pi}^{\mDP}_{t}=c^\dagger_{it}=\overline{c}_{it}$. Thus \eqref{eq:profitEQ} holds.
 \end{itemize}

\underline{Step 2.} We prove   $\hat{ \gbf}_i(\cbf_i+\boldsymbol{\epsilon}) \le \hat{ \gbf}_i(\cbf_i),    \boldsymbol{\epsilon}>0$ by the Monotone Comparative Statics Theorem of Topkis \cite[Theorem 6.1]{Topkis78}. In particular, since the feasible set of \eqref{eq:iniviG}  is a sublattice (Step 2.1), the objective of \eqref{eq:iniviG}  is supermodular in $\gbf_i$ and has decreasing differences in $(\mathbf{g}_i, \mathbf{c}_i)$ (Step 2.2), the argmax  optimal solution  $\hat{ \gbf}_i(\cbf_i)$ is nonincreasing in $\cbf_i$ under the componentwise order.

\underline{Step 2.1.} We prove that the constraint set of \eqref{eq:iniviG} is a lattice.  The box capacity constraints $\underline{r}^*_{it}\le g_{it}\le \overline{g}_i -\overline{r}^*_{it}$ and the initial ramping constraints $-r^{\D}_{i}\le g_{it'}-g_{i(t'-1)} \le r^{\U}_{i}$ are a product of intervals — a sublattice by inspection.

To prove ramping constraint $-\underline{r}^*_{it}\le g_{i(t+1)}-g_{it} \le \overline{r}^*_{it}$ forms a sublattice, denote two vectors $\gbf_i'$ and $\gbf_i''$ satisfying the ramping constraint. Let the $\vee$ (join) and $\wedge$ (meet) operations be defined by the element-wise max and min operations, respectively. If $g_{i(t+1)}' \ge g_{i(t+1)}''$, 
$$(g_{i(t+1)}' \vee g_{i(t+1)}'') \le \overline{r}^*_{it}+g_{it}'\le \overline{r}^*_{it}+(g_{it}' \vee g_{it}'')$$ 
$$\Rightarrow (g_{i(t+1)}' \vee g_{i(t+1)}'')  -(g_{it}' \vee g_{it}'')\le \overline{r}^*_{it}.$$ 
The same argument holds if $g_{i(t+1)}' \le  g_{i(t+1)}''$. Similarly, we can prove that 
$$-\underline{r}^*_{it}\le (g_{i(t+1)}' \vee g_{i(t+1)}'')  -(g_{it}' \vee g_{it}'')\le \overline{r}^*_{it}$$  
$$-\underline{r}^*_{it}\le (g_{i(t+1)}' \wedge g_{i(t+1)}'')  -(g_{it}' \wedge g_{it}'')\le \overline{r}^*_{it},$$ which satisfy the definition of a sublattice.

\underline{Step 2.2.} The objective of \eqref{eq:iniviG}  is (super)modular in $\mathbf{g}_i$ and has decreasing differences in $(\mathbf{g}_i, \mathbf{c}_i)$, because the objective $$f:=\sum_{t\in {\cal T}_{t'}}(\lambda_t^*  - c_{it}) g_{it}$$ is separable in the components of $\mathbf{g}_i$, so all cross-partials $$\frac{\partial^2 f}{\partial g_{it}\partial g_{is}}  = 0, t \neq s.$$ A separable function is modular, hence supermodular on any sublattice. Meanwhile, we have 
$$\frac{\partial^2 f}{\partial g_{it}\partial c_{it}}  = -1<0, \quad \frac{\partial^2 f}{\partial g_{it}\partial c_{is}}  =  0, t \neq s.$$ This implies that $f$ has decreasing differences in $(\mathbf{g}_i, \mathbf{c}_i)$.

\underline{Step 3 (Summary).} We compute the profit change of generator $i$ when increase bids from $\cbf_{i}$ to $\cbf_{i}+\boldsymbol{\epsilon}$ with constant $\boldsymbol{\epsilon}>0$. At time $t$,  we have
\beq\label{eq:Qdiff}
\begin{array}{lrl}
Q_{i}(\cbf_{i})-Q_{i}(\cbf_{i}+\boldsymbol{\epsilon})&\overset{(a)}{=} &\sum_{t\in {\cal T}_{t'}}  \big((\overline{c}_{it} -  c^\dagger_{it}) \hat{ g}_{it}(\cbf_{i}) \\
&&~~~~~~~~-(\overline{c}_{it} -  c^\dagger_{it})\hat{ g}_{it}(\cbf_{i}+\boldsymbol{\epsilon})\big) \\
&\overset{(b)}{\geq}& 0.
\end{array}
\eeq   
Here, (a) comes from  \eqref{eq:profitEQ} in Step 1. (b) relies Step 2, indicating $$\hat{g}_{it}(\cbf_{i})-  \hat{g}_{it}(\cbf_{i}+\boldsymbol{\epsilon})\geq 0, $$ and $\overline{c}_{it} \geq c_{it}^\dagger$ by definition.  Equation \eqref{eq:Qdiff} shows that  there is no incentive for a price-taking generator to bid higher than its true cost because the generator's profit $Q_i$ from \eqref{eq:profitQ}  is non increasing when the bid increases.  Therefore,  $\cbf_i=\cbf_i^\dagger$ maximizes profit of generator $i$ over ${\cal C}_i:=\{\cbf: \cbf^\dagger_{i} \le \cbf \le \overline{\cbf}_i\}$.  \hfill$\square$

\subsection{Additional results for simulations in the main text}\label{sec:AddSim}
%results reported in the main %text.

\subsubsection{Implementation Details}
\label{app:implementation}

{ 
We supplement the forecasting method in Sec.~II-D and the simulation settings in Sec.~IV-A with the following implementation details. The predictor's means and covariances are re-estimated at each forecast origin using a rolling 30-day historical data window. We assume that the binding-interval forecast is perfect, replacing the forecasted value with the realized net demand, while the advisory forecasts are retained.

The affine transformations are $d^{\mathrm{sim}}=0.022646d^{\mathrm{raw}}+71.91$ for CAISO and $d^{\mathrm{sim}}=0.015488d^{\mathrm{raw}}-152.08$ for ERCOT. This affine transformation is also applied to the forecasted values, whereas the uncertainty margins and FRP requirements are multiplied by the corresponding scaling factor. For ERCOT, days containing 15-minute net-demand changes more than five standard deviations from the mean change are excluded, and complete valid days are retained.  

The linear programs are implemented in Pyomo~6.9.0 and solved using Gurobi~12.0.1 with default numerical tolerances on a 64-bit Qualcomm ARMv8 processor with 16~GB RAM.

\subsubsection{Prediction-interval coverage}
Table~\ref{tab:forecast_coverage} reports the empirical out-of-sample coverage of the nominal 95\% prediction intervals used to construct the FRP requirements. Residual diagnostics indicate heavier tails and greater out-of-sample error variability than predicted, particularly during steep net-demand ramps, like the ones observed in CAISO's dataset during mornings and sunsets. Coverage falls below the nominal level during these ramps, whereas flatter periods generally exhibit coverage above 95\%. These findings are consistent with the overall undercoverage and show that the nominal 95\% level is a design target rather than an empirical guarantee.} %ndicating overly conservative intervals under those conditions. Thus, the margins do not fully capture how forecast uncertainty varies across operating conditions.

\begin{table}[t]
\centering
{ 
\caption{\scriptsize Empirical Coverage of Nominal 95\%
Prediction Intervals}
\label{tab:forecast_coverage}
\begin{tabular}{ccc}
\hline
Horizon & CAISO & ERCOT \\
\hline
15 min & 92.19\% & 93.87\% \\
30 min & 92.28\% & 93.82\% \\
45 min & 92.69\% & 93.79\% \\
60 min & 93.02\% & 93.60\% \\
\hline
\end{tabular}
}
\end{table}

\subsubsection{Interval BCR/MWP and Daily BCR/MWP}

\begin{table*}[htbp]
    \centering
\caption{Comparison of Daily-Settled Generator Profit across Ramping Scenarios and Pricing Methods (Unit: \$)}
    \label{tab:daily_profit_comparison}
    
    % --- Left Table: CAISO ---
    \begin{subtable}{0.48\textwidth}
        \centering
        \caption{CAISO Data}
        \label{tab:profit_caiso_appendix}
        \resizebox{\textwidth}{!}{%
        \begin{tabular}{l ccc ccc}
            \toprule
            & \multicolumn{3}{c}{\textbf{Single}} & \multicolumn{3}{c}{\textbf{Multi}} \\
            \cmidrule(lr){2-4} \cmidrule(lr){5-7}
            \textbf{\begin{tabular}{@{}l@{}}Ramp \\ Scenario\end{tabular}} & LMP & MTLMP & MDCP & LMP & MTLMP & MDCP \\
            \midrule
            S1 & 23816 & 85278 & 104498         & 27667 & 96179 & \textbf{107004} \\
            S3 & 55633 & 64821 & \textbf{67992} & 42430 & 57772 & 57446 \\
            S5 & 24126 & 32316 & \textbf{33629} & 26321 & 32372 & 33323 \\
            S8 & 30268 & 30268 & \textbf{30436} & 30256 & 30268 & \textbf{30436} \\
            \bottomrule
        \end{tabular}%
        }
    \end{subtable}%
    \hfill
    % --- Right Table: ERCOT ---
    \begin{subtable}{0.48\textwidth}
        \centering
        \caption{ERCOT Data}
        \label{tab:profit_ercot_appendix}
        \resizebox{\textwidth}{!}{%
        \begin{tabular}{l ccc ccc}
            \toprule
            & \multicolumn{3}{c}{\textbf{Single}} & \multicolumn{3}{c}{\textbf{Multi}} \\
            \cmidrule(lr){2-4} \cmidrule(lr){5-7}
            \textbf{\begin{tabular}{@{}l@{}}Ramp \\ Scenario\end{tabular}} & LMP & MTLMP & MDCP & LMP & MTLMP & MDCP \\
            \midrule
            S1 & 12963 & 20846 & \textbf{45800} & 13451 & 23080 & 44521 \\
            S3 & 15694 & 17353 & 17405          & 15294 & \textbf{17626} & 15903 \\
            S5 & 8426  & 8426  & \textbf{8471}  & 8426  & 8426  & \textbf{8471}  \\
            S8 & 8343  & 8343  & \textbf{8343}  & 8343  & 8343  & \textbf{8343}  \\
            \bottomrule
        \end{tabular}%
        }
    \end{subtable}
\end{table*}

\begin{table*}[htbp]
    \centering
    \caption{Comparison of Make-Whole Payments (MWP) and Percentage of Total Cost across Ramping Scenarios (Daily Settlement) (Unit: \$)}
    \label{tab:daily_mwp_comparison} % Changed label to avoid duplicate with main text
    
    % --- Left Table: CAISO ---
    \begin{subtable}{0.48\textwidth}
        \centering
        \caption{CAISO Data}
        \label{tab:mwp_caiso_appendix}
        \resizebox{\textwidth}{!}{%
        \begin{tabular}{l ccc ccc}
            \toprule
            & \multicolumn{3}{c}{\textbf{Single}} & \multicolumn{3}{c}{\textbf{Multi}} \\
            \cmidrule(lr){2-4} \cmidrule(lr){5-7}
            \textbf{\begin{tabular}{@{}l@{}}Ramp \\ Scenario\end{tabular}} & LMP & MTLMP & MDCP & LMP & MTLMP & MDCP \\
            \midrule
            S1 & 48790 (20.1\%) & 82 (0.0\%) & \textbf{0} (0.0\%) & 42551 (16.9\%) & 30 (0.0\%) & \textbf{0} (0.0\%) \\
            S3 & 241 (0.1\%)    & 13 (0.0\%) & \textbf{0} (0.0\%) & 621 (0.2\%)    & 29 (0.0\%) & \textbf{0} (0.0\%) \\
            S5 & 980 (0.3\%)    & 34 (0.0\%) & \textbf{0} (0.0\%) & 579 (0.2\%)    & 27 (0.0\%) & \textbf{0} (0.0\%) \\
            S8 & 2 (0.0\%)      & 2 (0.0\%)  & \textbf{0} (0.0\%) & 2 (0.0\%)      & 2 (0.0\%)  & \textbf{0} (0.0\%) \\
            \bottomrule
        \end{tabular}%
        }
    \end{subtable}%
    \hfill
    % --- Right Table: ERCOT ---
    \begin{subtable}{0.48\textwidth}
        \centering
        \caption{ERCOT Data}
        \label{tab:mwp_ercot_appendix}
        \resizebox{\textwidth}{!}{%
        \begin{tabular}{l ccc ccc}
            \toprule
            & \multicolumn{3}{c}{\textbf{Single}} & \multicolumn{3}{c}{\textbf{Multi}} \\
            \cmidrule(lr){2-4} \cmidrule(lr){5-7}
            \textbf{\begin{tabular}{@{}l@{}}Ramp \\ Scenario\end{tabular}} & LMP & MTLMP & MDCP & LMP & MTLMP & MDCP \\
            \midrule
            S1 & 8347 (3.7\%) & 240 (0.1\%) & \textbf{0} (0.0\%) & 7361 (3.2\%) & 207 (0.1\%) & \textbf{0} (0.0\%) \\
            S3 & 3 (0.0\%)    & 1 (0.0\%)   & \textbf{0} (0.0\%) & 8 (0.0\%)    & 1 (0.0\%)   & \textbf{0} (0.0\%) \\
            S5 & 1 (0.0\%)    & 1 (0.0\%)   & \textbf{0} (0.0\%) & 1 (0.0\%)    & 1 (0.0\%)   & \textbf{0} (0.0\%) \\
            S8 & 0 (0.0\%)    & 0 (0.0\%)   & \textbf{0} (0.0\%) & 0 (0.0\%)    & 0 (0.0\%)   & \textbf{0} (0.0\%) \\
            \bottomrule
        \end{tabular}%
        }
    \end{subtable}
\end{table*}

\begin{table*}[htbp]
    \centering
\caption{Comparison of Daily-Settled Demand Payment across Ramping Scenarios and Pricing Methods (Unit: \$)}
    \label{tab:demand_payment_appendix}
    
    % --- Left Table: CAISO ---
    \begin{subtable}{0.48\textwidth}
        \centering
        \caption{CAISO Data}
        \label{tab:dempay_app_caiso}
        \resizebox{\textwidth}{!}{%
        \begin{tabular}{l ccc ccc}
            \toprule
            & \multicolumn{3}{c}{\textbf{Single}} & \multicolumn{3}{c}{\textbf{Multi}} \\
            \cmidrule(lr){2-4} \cmidrule(lr){5-7}
            \textbf{\begin{tabular}{@{}l@{}}Ramp \\ Scenario\end{tabular}} & LMP & MTLMP & MDCP & LMP & MTLMP & MDCP \\
            \midrule
            S1 & \textbf{306413} & 367874 & 387095 & 309662          & 378174 & 388999 \\
            S3 & 331998          & 341187 & 344358 & \textbf{318336} & 333678 & 333352 \\
            S5 & \textbf{299061} & 307251 & 308564 & 301265          & 307316 & 308266 \\
            S8 & 305092          & 305092 & 305259 & \textbf{305079} & 305092 & 305259 \\
            \bottomrule
        \end{tabular}%
        }
    \end{subtable}%
    \hfill
    % --- Right Table: ERCOT ---
    \begin{subtable}{0.48\textwidth}
        \centering
        \caption{ERCOT Data}
        \label{tab:dempay_app_ercot}
        \resizebox{\textwidth}{!}{%
        \begin{tabular}{l ccc ccc}
            \toprule
            & \multicolumn{3}{c}{\textbf{Single}} & \multicolumn{3}{c}{\textbf{Multi}} \\
            \cmidrule(lr){2-4} \cmidrule(lr){5-7}
            \textbf{\begin{tabular}{@{}l@{}}Ramp \\ Scenario\end{tabular}} & LMP & MTLMP & MDCP & LMP & MTLMP & MDCP \\
            \midrule
            S1 & \textbf{240686} & 248570          & 273523          & 241139          & 250768          & 272210 \\
            S3 & 242930          & 244589          & 244641          & \textbf{242467} & 244799          & 243076 \\
            S5 & \textbf{235402} & \textbf{235402} & 235446          & \textbf{235402} & \textbf{235402} & 235446 \\
            S8 & \textbf{235312} & \textbf{235312} & \textbf{235312} & \textbf{235312} & \textbf{235312} & \textbf{235312} \\
            \bottomrule
        \end{tabular}%
        }
    \end{subtable}
\end{table*}

For completeness, we present the performance metrics evaluated on a daily MWP in Tables~\ref{tab:daily_profit_comparison}--\ref{tab:demand_payment_appendix}. Comparing these tables with the interval-based MWP presented in Section~IV-C, we confirm that calculating MWP on a daily basis reduces the total uplift payments compared to the interval-based calculation, as the daily formulation allows for netting between profitable and unprofitable intervals. Consequently, while MDCP and MTLMP still minimize uplifts, the relative magnitude of the BCR problem is smaller under daily settlement than under interval-by-interval settlement.

Regarding generator profits under LMP in the highly constrained scenarios (S1 and S3), as illustrated in Table~\ref{tab:daily_profit_comparison}: in these tight ramping scenarios, LMPs frequently drop or become negative during severe ramp-down events, causing the daily energy market revenues for certain units to fall below their total daily generation costs. When a generator operates at a daily loss in the energy market, the daily MWP perfectly offsets this exact shortfall to ensure bid cost recovery. Because the MWP effectively zeros out any net energy losses, these generators break even on their energy provision. For these generators that receive MWP, their total net profit is driven exclusively by their FRP payments. This boundary condition explains the profit values observed under LMP in scenarios S1 and S3. In contrast, under MDCP and MTLMP, the uniform in-market prices are inherently elevated to maintain positive net energy revenues in addition to the ramping compensation, resulting in the higher overall generator profits observed in the table when compared to LMP. Furthermore, because the price increase naturally offsets any potential energy losses from other moments of the day the magnitude of the total generator profits and demand payments under MDCP and MTLMP make the results very similar to the interval-based results reported in the main text.

\subsubsection{Interval-Level Price Volatility}
\begin{table}[t]
\caption{\scriptsize Interval-Level Price Standard Deviation in
Scenario S3 (\$/MWh)}
\label{tab:overall_price_std}
\centering
\scriptsize
\setlength{\tabcolsep}{4pt}
\begin{tabular}{ccccc}
\hline
Dataset & Dispatch & LMP & MDCP & MTLMP \\
\hline
CAISO & Single & 12.00 & 9.06 & 8.85 \\
CAISO & Multi  & 10.54 & 7.06 & 7.09 \\
ERCOT & Single &  9.92 & 10.04 & 10.02 \\
ERCOT & Multi  &  9.27 &  8.98 &  9.29 \\
\hline
\end{tabular}
\end{table}

\begin{figure}
    \centering
    \includegraphics[width=0.8\linewidth]{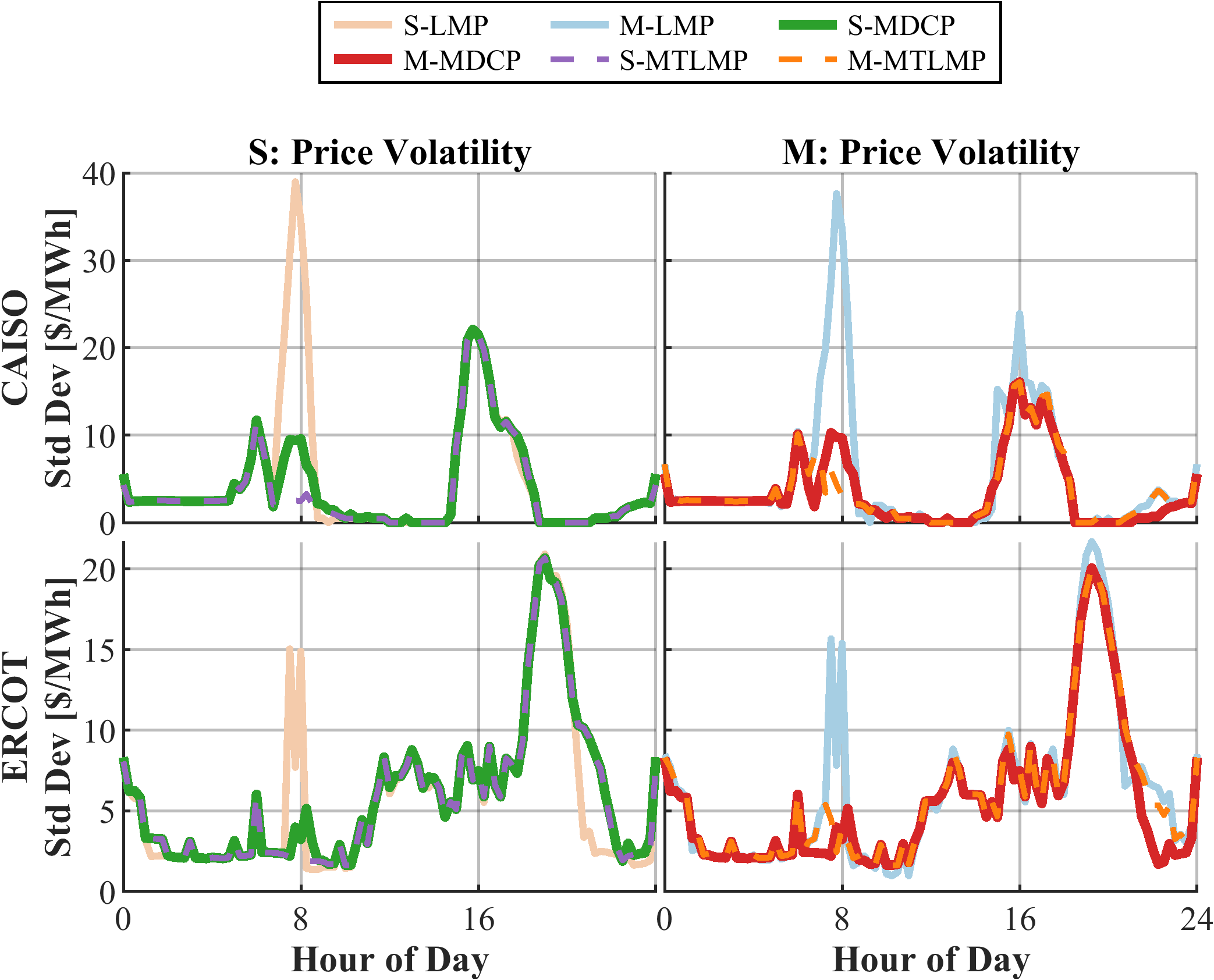}
\caption{\scriptsize Interval-level price volatility in scenario S3. Each curve reports the standard deviation of the binding-interval price at each 15-minute time of day. The left and right columns correspond to single- and multi-interval dispatch, respectively, while the upper and lower rows show the CAISO and ERCOT results. The associated mean price profiles are reported in Fig.~3.}
    \label{fig:price_variability}
    \vspace{-0.3cm}
\end{figure}

Fig.~\ref{fig:price_variability} reports the sample standard deviation of the binding-interval price across the 100 simulated days at each 15-minute time of day. Table ~\ref{tab:overall_price_std} reports the sample standard deviation across all binding-interval price observations. For CAISO, LMP exhibits the greatest variability, with overall standard deviations of $12.00$ and $10.54$ \$/MWh under single- and multi-interval dispatch, respectively. MDCP and MTLMP reduce these values, primarily by avoiding the extreme morning ramp-down price fluctuations observed under LMP. For ERCOT, the overall standard deviations are similar across all three mechanisms. Thus, the volatility reduction associated with MDCP and MTLMP is most pronounced for the more extreme CAISO ramping conditions and is not universal across datasets.

\subsubsection{Sensitivity to Forecast Uncertainty}
\label{subsec:sensitivity}

We test whether the relative performance of single and multi-interval dispatch is sensitive to forecast-error magnitude and distribution. To clarify the connection with the notation in the main text, let $r_t := d_t - d_{t-96}$ denote the net-demand residual, defined as the difference between the realized net demand $d_t$ and the 24-hour daily persistence profile $d_{t-96}$. We fit an autoregressive model of order 16, AR(16), by
\begin{equation}
 r_t=\phi_0+\sum_{k=1}^{16}\phi_k r_{t-k}+\varepsilon_t.
 \label{eq:ar16}
\end{equation}
The AR(16) parameters are estimated by ordinary least squares using the 15-minute residual observations preceding November 15, 2024. The net demand forecast $\hat{d}_t$ is then constructed from this model as $\hat{d}_t = d_{t-96} + (\phi_0 + \sum_{k=1}^{16}\phi_k r_{t-k})$. For each uncertainty case, we generate $N=200$ conditional net-demand trajectories with innovations
\begin{equation}
 \varepsilon_t^{(c)} =  \alpha_c\widehat{\sigma}_{\varepsilon}z_t^{(c)}, \qquad  z_t^{(c)} \sim
 \begin{cases}
  \mathcal{N}(0,1), & \text{Gaussian cases},\\
  \sqrt{3/5}\,T_5, & \text{Student-$t$ cases},
 \end{cases} \nonumber
\end{equation}
where $\widehat{\sigma}_{\varepsilon}$ is the estimated innovation standard deviation and $T_5$ has five degrees of freedom. The factor $\sqrt{3/5}$ gives the Student-$t$ innovation unit variance, allowing the effect of heavy tails to be separated from that of variance. We set $\alpha_c=1$ in the two base cases and $\alpha_c=2$ in the Gaussian-extreme and Student-$t$-combined cases; hence, the latter cases have four times the base innovation variance. Each evaluated trajectory contains 96 consecutive 15-minute intervals. The 95\% forecast-error bounds are obtained from the empirical 2.5th and 97.5th percentiles of $100{,}000$ four-step innovation paths propagated through the AR dynamics.

For ramp scenario $s$ and uncertainty case $c$, the sample-average daily cost difference is
\begin{equation}
 \widehat{\Delta C}_{s,c}
 =
 \frac{1}{N}\sum_{j=1}^{N}
 \left(C_{j,s,c}^{\mathrm{MI}}-C_{j,s,c}^{\mathrm{SI}}\right),
 \label{eq:cost_diff}
\end{equation}
where each daily cost includes generation and scarcity penalties. Negative values indicate that multi-interval (MI) dispatch is less expensive than single-interval (SI) dispatch.

\begin{table}[t]
\caption{\scriptsize Sample-Average Daily Cost Difference: MI Minus SI}
\label{tab:uncertainty_cost_difference}
\centering
\scriptsize
\setlength{\tabcolsep}{3pt}
\renewcommand{\arraystretch}{1.1}
\begin{tabular}{crrrr}
\hline
\textbf{Ramp} &
\shortstack{\textbf{Gaussian}\\\textbf{Base}} &
\shortstack{\textbf{Gaussian}\\\textbf{Extreme}} &
\shortstack{\textbf{Student-$t$}\\\textbf{Base}} &
\shortstack{\textbf{Student-$t$}\\\textbf{Combined}} \\
\hline
S1  & $-19{,}330.5$ & $-20{,}391.9$ & $-17{,}320.8$ & $-17{,}168.1$ \\
S5  & $-262.6$      & $-1{,}272.0$  & $-322.8$      & $-1{,}246.1$  \\
S8  & $0.0$         & $-0.3$        & $0.0$         & $-3.6$        \\
S10 & $0.0$         & $0.0$         & $0.0$         & $0.0$         \\
\hline
\multicolumn{5}{c}{\footnotesize Cost difference (\$/day)}\\
\end{tabular}
\end{table}

Table~\ref{tab:uncertainty_cost_difference} shows that the
sample-average cost difference is nonpositive in every tested configuration. MI reduces daily cost by approximately \$17,200--\$20,400 in S1, but the savings fall below \$1,300 in S5 and become negligible in S8 and S10. Across these cases, ramp availability has a substantially larger effect than the innovation distribution. Multi-interval dispatch saves costs when flexibility is scarce, whereas the two dispatch models perform nearly identically when sufficient flexibility is available. Multi-interval co-optimization outperforms single-interval dispatch when forecasts are accurate and ramping is tightly constrained, but this advantage diminishes under high forecast uncertainty and moderate ramping needs. 
 
\vspace{-0.2cm}
\subsection{Toy examples}\label{sec.toyEX}
In this section, we present 3-generator single-bus toy examples to illustrate the cause of MWP under LMP. The settings and results are summarized in Tables \ref{tb:upComplete} and \ref{tb:downComplete}. Generators G1, G2, and G3 have identical capacity limits of 500 MW, with ramp limits of 50, 50, and 10 MW/h, respectively. Their marginal costs are \$25, \$30, and \$50 /MWh. A penalty of \$80/MW is imposed on ramping slack, penalizing any inability to satisfy the ramping requirement. The detailed mathematical formulation is provided in (1). All capacities are in MW, and all prices and penalties are in \$/MWh.% Detailed ramping and multi-interval results are in the   appendix.

We present the dispatch and pricing results for both energy and ramping here. To focus on the key intuition, we construct cases where both single- and multi-interval dispatches produce the same dispatch result. We do hourly dispatch in this toy example. The single-interval dispatch determines results only for $t=1$. The multi-interval dispatch includes $t=1$ and $t=2$, but only $t=1$ is binding and implemented. $t=0$ represents the initial interval with given generation levels. For simplicity, in the table, we use ``New" to represent new uniform pricing methods proposed in this paper, which are MDCP and MTLMP. We denote single-interval LMP as S-LMP and multi-interval LMP as M-LMP. %The unit for dispatch is $MW$, for price is \$/MWh, for profit/payment is \$.

% To focus on the influence of pricing, we construct cases where both single- and multi-interval dispatches produce the same dispatch result. 
 % In Tables \ref{tb:I} and \ref{tb:II}, we denote single-interval LMP as S-LMP and multi-interval LMP as M-LMP, while MDCP and MTLMP results are presented jointly since they coincide in this case. More  comparisons between single- and multi-interval pricing and between MDCP and MTLMP are  in Sec.\ref{sec:SOAccessRight}.

%First we can observe the under compensation issue for both M-LMP and S-LMP. Ramp up toy example here only shows under compensation for M-LMP, ramp down show overcompensation for S-LMP and M-LMP. MWP will be used to fill in the negative profit part. Demand energy payment is computed by price payment in market and MWP paid by demand out of market.%We compare three pricing methods (LMP, ) in both

In the {\em ramp-up case} (Table \ref{tb:upComplete}), the demand forecast is 445 MW at $t=1$. For multi-interval dispatch the demand forecast is $\hat{\dbf}=(445, 555) $ MW. The ramp-up requirement at $t=1$ is 135 MW, which exceeds the available ramping capability and therefore requires 25 MW of ramping reserve, priced at \$80 /MW.  Under M-LMP, G1 has under compensation because the generator profit is \$ -6300 and, similarly, G2 experiences under-compensation \$ 500, meaning that their total payment under M-LMP is insufficient to cover their generation costs—necessitating MWP. Ramping payment cannot remove under-compensation and MWP, e.g. G1 receives negative profit even after accounting revenues from both   energy and ramping payments.%\footnote{Ramping payment details are shown in the  appendix.}

In the {\em ramp-down case} (Table \ref{tb:downComplete}),\footnote{S-LMP has under compensation because of the initial binding ramping constraints. There are also some research discussing reformulate the initial ramping as capacity limits and remove the under compensation for the non-scheduling part. This part is ignored in our paper here.} the demand forecast is 485 MW. For multi-interval dispatch the demand forecast is $\hat{\dbf}=(485, 450) $ MW. The ramp down requirement is set to be zero. Both S-LMP and M-LMP result in under-compensation for G2, again requiring MWP to ensure cost recovery.

In summary, under S-LMP and M-LMP, under-compensation issues persist, while MDCP/MTLMP eliminates discriminatory MWP and provides a consistent cost-reflective signal. Furthermore, both generator profits and total demand payments\footnote{Across both cases, we compute the demand energy payment as the sum of the market price payment and out of market MWP.} under MDCP/MTLMP are higher than under LMP, as all participants are compensated based on the highest scheduled marginal cost.

 \begin{table}[ht] 
\centering 
\caption{\scriptsize Ramp up case: dispatch and pricing results at $t=1$ are binding with initial generation $\mathbf{g}_0=(370, 50, 0)^\intercal$. The unit for dispatch is MW, for price is \$/MWh, for profit/payment is \$. The columns reporting generator profit and demand payment correspond only to the binding interval $t=1$.}\label{tb:upComplete}
\renewcommand{\arraystretch}{1.2}
\begin{tabular}{l|c|c|c|c|c|c}
\toprule
 & \multicolumn{3}{c|}{dispatch/pricing results}  & \multicolumn{3}{c}{generator profit/demand payment}  \\\midrule
\textbf{G} & \textbf{t=0} & \textbf{t=1} & \textbf{t=2} & \textbf{S-LMP} & \textbf{M-LMP} & \textbf{New} \\
\midrule
G1 & 370 & 420 & 470 & 2100 & -6300 & 2100 \\
G2 & 50 & 25 & 75 & 0 & -500 & 0 \\
G3 & 0 & 0 & 9 & 0 & 0 & 0 \\
D  & 420 & 445 & 554 & 13350 & 11250 & 13350 \\
\midrule
S-LMP  & - & 30 & - & - & - & - \\
M-LMP  & - & 10 & 50 & - & - & - \\
New & - & 30 & 50 & - & - & - \\
% MTLMP  & - & 30 & 50 & - & - & - \\
% S-MTLMP  & - & 30 & - & - & - & - \\
\midrule
R1 & 50 & 50 & - &\multicolumn{3}{c}{4000} \\
R2 & 50 & 50 & - & \multicolumn{3}{c}{4000} \\
R3 & 10 & 10 & - & \multicolumn{3}{c}{800} \\
$\overline{\omega}$ & 0 & 135 & - & - & - & - \\
Slack & 0 & 25 & - & - & - & - \\
\midrule
$\pi^{\U}$ & - & 80 & - & - & - & - \\
\bottomrule
\end{tabular}
%\vspace{-0.4cm}
\end{table}

\begin{table}[ht]
\centering
\caption{\scriptsize Ramp down case: dispatch and pricing results at $t=1$ are binding with initial generation $\mathbf{g}_0=(470, 60, 0)^\intercal$. The unit for dispatch is MW, for price is \$/MWh, for profit/payment is \$. The columns reporting generator profit and demand payment correspond only to the binding interval $t=1$.}\label{tb:downComplete}
\renewcommand{\arraystretch}{1.2}
\begin{tabular}{l|c|c|c|c|c|c}
\toprule
 & \multicolumn{3}{c|}{dispatch and pricing results}  & \multicolumn{3}{c}{generator profit/demand payment}  \\\midrule
\textbf{G} & \textbf{t=0} & \textbf{t=1} & \textbf{t=2} & \textbf{S-LMP} & \textbf{M-LMP} & \textbf{New} \\
\midrule
G1 & 470 & 475 & 450 & 0 & 0 & 2375 \\
G2 & 60 & 10 & 0 & -50 & -50 & 0 \\
G3 & 0 & 0 & 0 & 0 & 0 & 0 \\
D  & 530 & 485 & 450 & 12175 & 12175 & 14550 \\
\midrule
S-LMP  & - & 25 & - & - & - & - \\
M-LMP  & - & 25 & 25 & - & - & -\\
New & - & 30 & 25 & - & - & - \\
% MTLMP  & - & 30 & - & - & - & - \\
% \midrule
% r1 & 50 & 50 & - &\multicolumn{3}{c}{1250} \\
% r2 & 50 & 2 & - & \multicolumn{3}{c}{1250} \\
% r3 & 0 & 0 & - & \multicolumn{3}{c}{50} \\
% $\hat{\omega}^{\D}$ & 0 & 52 & - & - & - & - \\
% Slack & 0 & 0 & - & - & - & - \\
% \midrule
% $\pi^{\D}$ & - &  0 & - & - & - & - \\
\bottomrule
\end{tabular}
%\vspace{-0.5cm}
\end{table}

\vspace{-0.2cm}
\subsection{Generalization: network congestion}

When consider power network constraints, the DC OPF model will be included in (1). That way, different buses will have different LMPs when there are network congestion. We here comment on extensions of MDCP and MTLMP to the case with network congestion. MTLMP can be naturally extended to the case with network congestion. We anticipate properties of MTLMP will stay the same when network constraints are considered. 

MDCP doesn't have a direct extension to consider network congestion. When there's network congestion, we develop the following pricing run to find the MDCP-like pricing solutions, named as uniform price with minimum uplift (UPMU). The goals is to find price adders $\pmb{\Delta}_t:=(\Delta_{j,t})_{j=1}^N$ for each node $j$ with small adjustment to LMP but minimum out of market MWP uplifts. The UPMU is defined by $\pibf^{\text{\tiny UPMU}}_{t}:= \pibf^{\text{\tiny LMP}}_{t} + \pmb{\Delta}_t$.
 \begin{subequations} \label{eq:pricingN}
\begin{align}
&\underset{\pmb{\Delta}_t:=(\Delta_{j,t})_{j=1}^M}{\rm minimize} &&  \alpha||\pmb{\Delta}_t||+\sum\limits_{j=1}^M{\cal M}_{j,t}(\pi^{\text{\tiny LMP}}_{j,t} + \Delta_{j,t}, g^{\R}_{j,t}) \nn\\
&&&~~~~~~~~~~~~~~~ +\beta\sum\limits_{j=1}^M (\pi^{\text{\tiny LMP}}_{j,t} + \Delta_{j,t})d_{j,t}\\
& {\rm subject~to} &&  \sum\limits_{j=1}^M \Delta_{j,t}(d_{j,t} - g^{\R}_{j,t})\geq 0, \forall j \in [M]. \label{eq:deltaMS0}
%& && \mbox{MWP}_{j,t} = 0,\label{eq:MWP0}\\
% & && 
%&&& \pibf_t \geq \pibf^{\text{\tiny LMP}}_{t}.
%& & |\pi_{a,t}-\pi_{b,t}| <= \alpha |\pi^{\text{\tiny LMP}}_{a,t}-\pi^{\text{\tiny LMP}}_{b,t}|, \forall a, b \in {\cal M}.
\end{align}
\end{subequations}
Here, we denote $d_{j,t}$  as the demand  at time $t$ on bus $j$. $[M]$ is the set including all $M$ buses in the transmission network. For simplicity, we assume one generator at each bus. Note that $\pi^{\text{\tiny UPMU}}_{j,t}$ is also a  nodal price as LMP, which is uniform for all resources at the same bus. 
$g^{\R}_{j,t}$ represents  total generation at bus $j$ from the binding interval of the rolling-window dispatch run optimization (1)-(2). 
  
%  \begin{subequations}\label{eq:pricingimprove}
% \begin{align}
% &\underset{\pibf_t:=(\pi_{j,t})}{\rm minimize} &&  \sum\limits_{j\in {\cal M}}  ||\pibf_t - \pibf^{\text{\tiny LMP}}_{t}|| \\& subject~to && \forall j \in [N],\\& && \mbox{MWP}_{j,t} = 0,\\
% & && \sum\limits_{j\in {\cal M}} (\pi_{j,t}-\pi^{\text{\tiny LMP}}_{j,t})(d_{j,t} - g_{j,t})\geq 0,\\\label{eq:deltaMS>=0}
% &&& \pibf_t \geq \pibf^{\text{\tiny LMP}}_{t}.
% %& & |\pi_{a,t}-\pi_{b,t}| <= \alpha |\pi^{\text{\tiny LMP}}_{a,t}-\pi^{\text{\tiny LMP}}_{b,t}|, \forall a, b \in {\cal M}.
% \end{align}
% \end{subequations}

%  \begin{subequations}\label{eq:pricingimprove2}
% \begin{align}
% &\underset{\pibf_t:=(\pi_{j,t})}{\rm minimize} &&  \sum\limits_{j\in {\cal M}}  ||\pibf_t - \pibf^{\text{\tiny LMP}}_{t}|| \\& subject~to && \forall j \in [N],\\& && \mbox{MWP}_{j,t} = 0,\\
% & && \sum\limits_{j\in {\cal M}} \pi_{j,t}(d_{j,t} - g_{j,t})\geq 0.\label{eq:deltaMS>=0}
% % &&& \pibf_t \geq \pibf^{\text{\tiny LMP}}_{t}.
% %& & |\pi_{a,t}-\pi_{b,t}| <= \alpha |\pi^{\text{\tiny LMP}}_{a,t}-\pi^{\text{\tiny LMP}}_{b,t}|, \forall a, b \in {\cal M}.
% \end{align}
% \end{subequations}

 % We run this single interval pricing run \eqref{eq:pricingN} to solve for the real-time MDCP at time $t$.

Essentially, \eqref{eq:pricingN} is the pricing run optimization to compute the real time price adders for UPMU. This pricing run takes the dispatch results from (1) and directly enforce all properties we required to get the uniform price for the rolling-window dispatch at each bus. When minimizing uplifts MWP in the objective,  $\sum_{j=1}^M{\cal M}_{j,t}=0$ can be satisfied when price adders are high enough for all generators. By setting weights $\alpha, \beta \geq 0$ properly, UPMU solved from this optimization \eqref{eq:pricingN} (i) gives the optimizer MDCP when there is no network congestion, which recovers the closed-form solution in the main text; (ii) has the optimal solution equals LMP when there's no binding ramping constraints. This is supported by the objective minimizing price difference  $ ||\pmb{\Delta}_t||=||\pibf_t^{\text{\tiny UPMU}} - \pibf^{\text{\tiny LMP}}_{t}||$.

% This pricing run optimization is always feasible because zero price adder is one feasible solution. 
As for the merchandising surplus (MS), we can always satisfy the nonnegative MS constraint in \eqref{eq:deltaMS0} by increasing price adders. If $\sum_{j\in {\cal M}} \Delta_{j,t}(d_{j,t} - g_{j,t}) < 0$, find the bus with more demand\footnote{In a congested network, we can always find some buses with more demand and some buses with more generations.} and increase the price.

Financial transmission rights (FTR) are influenced by the nodal prices. We compute the FTR for each line by the branch power flow multiplied by the nodal price differences for all MDCP, MTLMP, UPMU, and LMP.  That way, we can always guarantee FTR payment equals MS. Although we have nonnegative MS guaranteed by \eqref{eq:deltaMS0}, MS under MDCP doesn't equal that under LMP. 

 {Our latest progress 
extends  MDCP to congested network  through an optimal pricing formulation that can preserve zero MWP and develops a closed-form extension and examines truthful-bidding properties~\cite{norambuenaguzman2026optimal}.}

%Also, when there's network congestions and binding ramping constriant simultaneously in the rolling-window dispatch, we no longer have congestion rent computed by the dual variable of network line capacity for congestion rent. We still have nodal price difference for FTR.

\subsection{IEEE RTS-24 Study}
\label{app:rts24_fleet}

To determine whether the conclusions of the main numerical study extend to a larger generation fleet, we repeated the experiment in the main text to the IEEE RTS-24 system from \cite{ordoudis2016updated}.
Generator capacities and marginal costs are taken directly from that reference. Transmission limits are assumed to be nonbinding, so energy and FRP prices are system-wide.  The 15-minute CAISO net-demand profile from the main study is rescaled affinely so that its minimum and maximum match the corresponding load levels of the RTS-24 profile in \cite{ordoudis2016updated}. The rolling-window dispatch, FRP requirement definitions, pricing mechanisms, and settlement calculations are unchanged.

To isolate the effect of ramping capability, we replace the ramp rates reported in the RTS-24 data with controlled scenario-specific ramping limits expressed as fractions of each generator’s capacity. We first partition the fleet into low, medium, and high cost groups:
\begin{align}
 \mathcal{G}^{\mathrm{L}}
 &=\{G_8,G_9,G_{10}\}, &
 \mathcal{G}^{\mathrm{M}}
 &=\{G_1,G_2,G_6,G_7,G_{11},G_{12}\}, \nonumber\\
 \mathcal{G}^{\mathrm{H}}
 &=\{G_3,G_4,G_5\}. \nonumber
 \label{eq:rts24_groups}
\end{align}
For generator $i$ in scenario $s$, the 15-minute ramp capability is
\begin{equation}
 R_{i,s}^{U}=R_{i,s}^{D}
 =\beta_{s,h(i)}\overline{g}_i, \nonumber
 \label{eq:rts24_ramp_limits}
\end{equation}
where $h(i)\in\{\mathrm{L},\mathrm{M},\mathrm{H}\}$ denotes the cost group that contains generator $i$, and $\beta_{s,h(i)}$ is the corresponding ramp-to-capacity ratio. Following the logic of the original experiment, ramping flexibility is first increased for the low-cost generators, followed by the medium and high-cost generators. In S7, all three groups reach a ramp-to-capacity ratio of 10\%; scenarios S8--S11 then increase the capability of the entire fleet uniformly.

\begin{table}[t]
\caption{Ramp-Capability Scenarios for the RTS-24 Generation Mix}
\label{tab:rts24_ramp_scenarios}
\centering
\scriptsize
\setlength{\tabcolsep}{3.2pt}
\renewcommand{\arraystretch}{0.95}
\begin{tabular}{ccccc}
\hline
Scenario & $\beta_{s,\mathrm{L}}$ & $\beta_{s,\mathrm{M}}$
& $\beta_{s,\mathrm{H}}$ & \shortstack{Total 15-min \\ ramp capability (MW)} \\
\hline
S1  & 4\%     & 3\%     & 3\%     & 112.25 \\
S2  & 6\%     & 3\%     & 3\%     & 134.25 \\
S3  & 10\%    & 3\%     & 3\%     & 178.25 \\
S4  & 10\%    & 6\%     & 3\%     & 216.47 \\
S5  & 10\%    & 10\%    & 3\%     & 267.43 \\
S6  & 10\%    & 10\%    & 6\%     & 297.46 \\
S7  & 10\%    & 10\%    & 10\%    & 337.50 \\
S8  & 15\%    & 15\%    & 15\%    & 506.25 \\
S9  & 20\%    & 20\%    & 20\%    & 675.00 \\
S10 & 50\%    & 50\%    & 50\%    & 1687.50 \\
S11 & 99.98\% & 99.98\% & 99.98\% & 3374.33 \\
\hline
\end{tabular}
\end{table}
The final column of Table~\ref{tab:rts24_ramp_scenarios}
reports the total generators ramp capability in either direction.
Table~\ref{tab:app_gen_profit} presents the average generator profits for the representative scenarios S1, S3, S5, S7, and S11. The omitted scenarios exhibit the same qualitative behavior. In S1, the average daily system-wide generator profit is $23.2\%$ higher under MDCP and $26.4\%$ higher under MTLMP than under LMP with MWP. As ramping capability is increased in the medium and high-cost groups, these profit differences decline to $17.8\%$ and $19.1\%$ in S5, and to $2.9\%$ and $3.6\%$ in S7. By S11, all three mechanisms produce the same profit, indicating that the results converge once the intertemporal ramp constraints are no longer binding.

{As Tables~\ref{tab:app_mwp} and \ref{tab:app_demand_payment} show, total demand payments closely tracked aggregate generator profits. In highly ramp-constrained scenarios such as S1, LMP still required discriminatory MWPs for bid-cost recovery, averaging 8.2\% of generators' total energy revenue. MDCP eliminated these interval-level MWPs, and MTLMP reduced them to negligible levels. Because MDCP and MTLMP cover the shortfall through the uniform price, which pays every generator the higher price, generator profits and demand payments were higher than under LMP with MWP. This comparison assumes truthful bidding and may not hold if generators bid strategically to collect MWPs (Sec.~III-E). In S1, the LMP MWPs of the IEEE RTS-24 system were at the low end of the 6.4--41.1\% range in the three-generator system of the main text, likely because the larger fleet provides more ramping capability, so binding ramp constraints occur less often.}

 MTLMP with MWPs produces slightly higher aggregated profit than MDCP in the most constrained RTS-24 scenarios. This difference reflects both generator diversity and the distinct cost-recovery guarantees of the two mechanisms. In the three-generator system, tight ramping constraints frequently brought the generator with the highest cost into the dispatch, increasing MDCP. With 12 smaller generators with similar variable costs, the scheduled movement can be distributed across more units, so the highest active marginal cost, and therefore MDCP, might not increase as sharply. At the same time, binding intertemporal constraints introduce temporal opportunity-cost components that can raise MTLMP. For a fixed dispatch, MDCP is the minimum uniform price that covers the bid-in costs of all dispatched generators, thus eliminating interval-level MWPs while limiting additional producer surplus within the class of zero-MWP uniform settlements. MTLMP does not guarantee interval-level bid-cost recovery, so residual MWPs may still be required. The higher aggregate profit under MTLMP with MWP therefore reflects the effect of intertemporal binding ramping constraints, rather than a strict ordering between MTLMP and MDCP.

\begin{table}[htbp]
\centering
\caption{Comparison of Average Generator Profit across Ramping Scenarios (Unit: \$)}
\label{tab:app_gen_profit}
\begin{tabular}{lccc}
\toprule
Ramp Scenario & LMP+MWP & MDCP & MTLMP+MWP \\
\midrule
S1 & 741048 & 912828 & \textbf{936366 }\\
S3 & 683606 & 867830 & \textbf{881660 }\\
S5 & 529145 & 623341 & \textbf{630093 }\\
S7 & 435281 & 447814 & \textbf{450770 }\\
S11 & \textbf{422892} & \textbf{422892 }&\textbf{ 422892} \\
\bottomrule
\end{tabular}
\end{table}

\begin{table}[htbp]
\centering
\caption{Comparison of Average MWP and Percentage Relative to Pure Energy Revenue (Unit: \$)}
\label{tab:app_mwp}
\begin{tabular}{lccc}
\toprule
Ramp Scenario & LMP & MDCP & MTLMP \\
\midrule
S1 & 87134 (8.2\%) & 0 (0.0\%) & 34 (0.0\%) \\
S3 & 26482 (2.5\%) & 0 (0.0\%) & 0 (0.0\%) \\
S5 & 3252 (0.3\%) & 0 (0.0\%) & 0 (0.0\%) \\
S7 & 965 (0.1\%) & 0 (0.0\%) & 0 (0.0\%) \\
S11 & 0 (0.0\%) & 0 (0.0\%) & 0 (0.0\%) \\
\bottomrule
\end{tabular}
\end{table}

\begin{table}[htbp]
\centering
\caption{Comparison of Average Demand Payment across Ramping Scenarios (Unit: \$)}
\label{tab:app_demand_payment}
\begin{tabular}{lccc}
\toprule
Ramp Scenario & LMP+MWP & MDCP & MTLMP+MWP \\
\midrule
S1 & \textbf{1170829} & 1342609 & 1366147 \\
S3 & \textbf{1113802} & 1298026 & 1311855 \\
S5 & \textbf{951127} & 1045323 & 1052074 \\
S7 & \textbf{856546} & 869078 & 872034 \\
S11 & \textbf{842869} &\textbf{ 842869} & \textbf{842869} \\
\bottomrule
\end{tabular}
\end{table}

\subsection{Detail derivations of MTLMP}\label{eq:MLMPderive}

To compute the TLMP of generator $i$ at time $t$, we fix its generation at the optimal value $g^*_{it}$ and all ramping awards at their optimal values $\rbf^*:=\{\overline{\rbf}^*_k,\underline{\rbf}^*_k\}_{k\in{\cal T}_t}$, which excludes the effect of the ramping products on the energy price. Let $\{g_{-it}\}$ denote all other dispatch variables in the window, including the initial levels $g_{j(t-1)}$. The minimum cost of these decisions, $V_{it}(g^*_{it},\rbf^*)$, is the optimal value of
\begin{equation}\label{eq:EDMTLMP}
\begin{aligned}
\underset{\gbf_{-it}}{\text{minimize}}\quad & \sum_{j\in [N]}\sum_{k\in \mathcal{T}_{t}} c_{jk} g_{jk} - c_{it} g^*_{it}\\[3pt]
\text{subject to}\quad & \sum_{j\in [N]} g_{jk} = \hat{d}_{k},\quad \forall k \in \mathcal{T}_{t}\setminus\{t\},\\
\lambda_t:~~ & \sum_{j\in [N]\setminus\{i\}} g_{jt} + g^*_{it} = \hat{d}_{t},\\
& -\underline{r}^*_{jk} \le g_{j(k+1)} - g_{jk} \le \overline{r}^*_{jk},\\
&  \forall j\in [N]\setminus\{i\},~\forall k \in \mathcal{T}_{t}\setminus\{t+W-1\},\\
(\underline{\gamma}_{i(t+1)},\overline{\gamma}_{i(t+1)}):~~ & -\underline{r}^*_{it} \le g_{i(t+1)} - g^*_{it} \le \overline{r}^*_{it},\\
(\underline{\gamma}_{it},\overline{\gamma}_{it}):~~ & -r^{\D}_{i} \le g^*_{it} - g_{i(t-1)} \le r^{\U}_{i},\\
& -r^{\D}_{j} \le g_{jt} - g_{j(t-1)} \le r^{\U}_{j},  \forall j \in [N]\setminus\{i\},\\
& \underline{r}^*_{jk} \le g_{jk} \le \overline{g}_j - \overline{r}^*_{jk},\\
& \qquad \forall j \in [N]\setminus\{i\},~\forall k \in \mathcal{T}_{t}\setminus\{t\},\\
& g_{j(t-1)} = x_j,\quad \forall j\in[N],
\end{aligned}
\end{equation}
where the last constraint sets the initial generation levels $x_j$ to the dispatch realized in the previous window. By the envelope theorem,
\beq
 \pi^{\TLMP}_{it}:= - \frac{\partial}{\partial g_{it}} V_{it}(g^*_{it}, \rbf^*)\\
 =\lambda^*_t-\underline  \gamma_{i(t+1)} ^*+\overline \gamma_{i(t+1)}^* +\underline \gamma_{it}^* - \overline \gamma_{it}^*,\nn
 \eeq
from which MTLMP follows by \eqref{eq:mtlmp}.

Problem \eqref{eq:EDMTLMP} may not have a unique optimal dual solution, and in special cases neither does \eqref{eq:ED}. Matching the KKT conditions of the two problems shows that the optimal dual variables of \eqref{eq:ED} are also optimal for \eqref{eq:EDMTLMP}, and every such dual solution satisfies the MTLMP properties derived in Sec.~III. When the dual is not unique, our simulations use the dual solution of \eqref{eq:ED} returned by Gurobi with  settings in Sec.IV-A.%; its results are based on the choosing method ...deterministic for a given model and parameter settings.%, so the reported prices are reproducible.

\end{document}